\documentclass[letterpaper]{article} 
\usepackage{aaai23}  
\usepackage{times}  
\usepackage{helvet}  
\usepackage{courier}  
\usepackage[hyphens]{url}  
\usepackage{graphicx} 
\usepackage{natbib}  
\usepackage{caption} 
\usepackage{algorithm}
\usepackage{algorithmic}

\usepackage{graphics}
\usepackage{multicol}
\usepackage{multirow}
\usepackage{xcolor}
\usepackage{amsmath}
\usepackage{amssymb}
\usepackage{amsthm}
\usepackage{enumitem}
\usepackage{bm}
\usepackage{booktabs}
\usepackage{units}
\definecolor{orange}{RGB}{255, 127, 0}

\newtheorem{theorem}{Theorem}
\theoremstyle{definition}
\newtheorem{definition}{Definition}
\newtheorem{proposition}{Proposition}
\newcommand{\modelname}{\texttt{PPGenCDR}}
\newcommand{\moduleAname}{\textsl{SPPG}} 
\newcommand{\moduleBname}{\textsl{RCDR}}
\usepackage{subfigure}
\newcommand{\nosection}[1]{\noindent\textbf{#1.}}

\usepackage{newfloat}
\usepackage{listings}
\DeclareCaptionStyle{ruled}{labelfont=normalfont,labelsep=colon,strut=off} 
\floatstyle{ruled}
\newfloat{listing}{tb}{lst}{}
\floatname{listing}{Listing}
\title{PPGenCDR: A Stable and Robust Framework for Privacy-Preserving Cross-Domain Recommendation}
\author{
    Xinting Liao\textsuperscript{\rm 1},
    Weiming Liu\textsuperscript{\rm 1},
    Xiaolin Zheng\textsuperscript{\rm 1},
    Binhui Yao\textsuperscript{\rm 2},
    Chaochao Chen\textsuperscript{\rm 1}\thanks{Chaochao Chen is the corresponding author.}\\
}
\affiliations{
    \textsuperscript{\rm 1}College of Computer Science and Technology, Zhejiang University, China\\
    \textsuperscript{\rm 2}Midea Group, Foshan, China\\
    \{xintingliao, 21831010, xlzheng, zjuccc\}@zju.edu.cn,
    tony.yao@midea.com\\

}

\usepackage{bibentry}

\begin{document}

\maketitle

\begin{abstract}
Privacy-preserving cross-domain recommendation (PPCDR) refers to preserving the privacy of users when transferring the knowledge from source domain to target domain for better performance, which is vital for the long-term development of recommender systems.
Existing work on cross-domain recommendation (CDR) reaches advanced and satisfying recommendation performance, but mostly neglects preserving privacy.
To fill this gap, we propose a privacy-preserving generative cross-domain recommendation (\modelname) framework for PPCDR.
\modelname~includes two main modules, i.e., \textit{stable privacy-preserving generator} module, and \textit{robust cross-domain recommendation} module.
Specifically, the former isolates data from different domains with a generative adversarial network (GAN) based model, which stably estimates the distribution of private data in the source domain with \'Renyi differential privacy (RDP) technique. 
Then the latter aims to robustly leverage the perturbed but effective knowledge from the source domain with the raw data in target domain to improve recommendation performance. 
Three key modules, i.e., \textit{(1) selective privacy preserver}, \textit{(2) GAN stabilizer}, and \textit{(3) robustness conductor}, guarantee the cost-effective trade-off between utility and privacy, the stability of GAN when using RDP, and the robustness of leveraging transferable knowledge accordingly.
The extensive empirical studies on Douban and Amazon datasets demonstrate that \modelname~significantly outperforms the state-of-the-art recommendation models while preserving privacy.

\end{abstract}

\section{Introduction}
In recent years, cross-domain recommendation (CDR) is boosting for its effectiveness in alleviating the sparsity problem by transferring informative knowledge across related domains~\cite{zhu2021cross}.
On the one hand, the user-item interaction data 
plays a vital role in enlightening the recommendation performance in CDR~\cite{zhu2021unified,chen2020towards,li2020ddtcdr}. 
On the other hand, the interaction data is closely related to user privacy.

\begin{figure}[t]
\centering
\includegraphics[width=1\columnwidth]{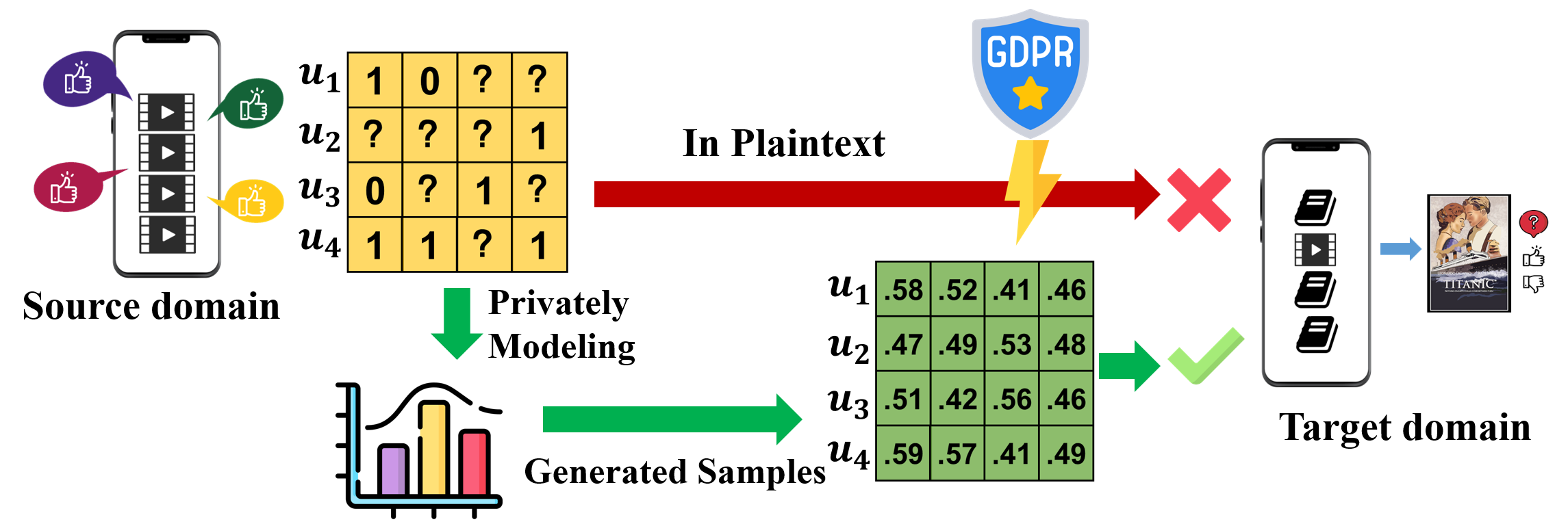} %
\caption{An example of the necessity of \modelname. 
}
\label{fig:motivation}
\end{figure}

Although existing work on CDR achieves advanced performance, most of them assume that it is accessible to transfer interaction data across domains in plaintext~\cite{chen2020towards,li2020ddtcdr}. 
In reality, different domains maintain different types of user-sensitive data, e.g., Amazon has rich user-commodity interaction data, while IMDb is rich in user-movie interaction data. 
As shown in Fig.~\ref{fig:motivation}, it assumes that directly transferring the knowledge without privacy-preserving of a source domain (movie) to boost the performance of a target domain (book) is accessible. 
However, constrained by legal regulations,
a domain may not be able to utilize the interaction data of other domains in plaintext.
Therefore, privacy-preserving CDR (PPCDR) is vital
for the long-term development of recommender systems. 

Recently, several work was proposed to preserve the privacy of recommender systems in two data storage locations, i.e., (1) \textit{the end-user device} that generates data, and (2) \textit{the central database} that legally collects data uploaded by the end-user devices.
The \textit{former} avoids publishing user data to the server by modeling the data locally, while unifying a model globally for maintaining performance~\cite{ammad2019federated}.
However, the efficiency of the former is limited by the frequent communications among local devices and  central server (if any).  
The \textit{latter} prevents privacy leakage from modeling data \cite{Liu2022Privacy-Preserving}, or transferring information among different central platforms~\cite{gao2019privacy,cui2021exploiting}.
Nevertheless, these methods do not devote to protecting the interaction data of users in CDR.
PriCDR~\cite{chen2022differential} directly applies differential privacy (DP) to interaction data in a source domain and publishes the perturbed data to a target domain. 
However, PriCDR \textit{causes  looser privacy budget bound compared with using \'Renyi DP (RDP)}~\cite{mironov2017renyi} and \textit{can not perform recommendation robustly}~\cite{gao2022self}.

To take advantage of interaction data, and bypass its disadvantage of leaking privacy, we devote to using \textit{privacy-preserving data publishing} in PPCDR,  
which is a popular technique to release data to public in the perturbed form~\cite{dwork2009complexity,fung2010privacy,beaulieu2019privacy}.
As Fig.~\ref{fig:motivation} depicts, we use privacy-preserving data publishing to isolate the raw data (the yellow matrix) in the source domain from target domain.
To reach this goal, we propose a framework based on generative adversarial network~(GAN), namely, privacy-preserving generative model-based CDR~(\modelname) in this work.
Firstly, we adversarially model the distribution of the raw data in the source domain with RDP.
Then we generate dense but fake ratings (the green matrix) to target domain.

Nevertheless,  three challenges make it non-trivial to implement~\modelname.
Firstly, \textbf{CH1}:  \textit{How to preserve privacy in the most economical way?}
In order to preserve privacy during modeling, DP clips and perturbs the gradients of every layer in the model~\cite{abadi2016deep}. 
This approach reduces the utility of the model due to adding a lot of noise, but gains insignificant benefits of privacy-preserving. 
To address \textbf{CH1}, \modelname~contains a \textbf{selective privacy preserver (SPP)}, which only applies RDP on the gradients of model layer that directly accesses the raw data. 
The overall privacy can be theoretically guaranteed by post-processing~\cite{dwork2014algorithmic} and chain rule.

Secondly, \textbf{CH2}: \textit{How to improve the stability of \modelname~when using RDP to preserve privacy?}
%
GAN-based model optimizes a minmax goal, which is difficult to converge under the game of generator and discriminator \cite{roth2017stabilizing}.
%
It worsens the process of modeling data when applying RDP to perturb the gradients of model in source domain. 
To tackle \textbf{CH2},
we design a \textbf{GAN stabilizer (GS)} to stabilize the objective of discriminator by an extra regularization derived from control theory.

Lastly, \textbf{CH3}: \textit{How to improve the robustness of CDR models in target domain at the lowest cost when leveraging perturbed data?}
There are several advanced CDR models in target domain that perform well in modeling source domain data in plaintext. 
But it requires a costly modification for these models to adapt to the perturbed data.
To resolve \textbf{CH3}, we leverage \textbf{robust conductor (RC)} as an extra and flexible plugin of CDR model.
Specifically, RC improves the model performance in target domain by disentangling each dimension of the redundant representations in a batch.
%

%
We summarize our contributions as follows: 
(1) We are the first to privately model the distribution of interaction data, and transfer it as the knowledge of source domain to improve the recommendation performance in target domain. 
(2) We propose a~\modelname~framework with three special modules, i.e., \textit{SPP}, \textit{GS}, and \textit{RC}, to stably model the raw data of source domain and robustly enhance the performance of target domain in a privacy-preserving way.
(3) We conduct extensive experiments to empirically prove the utility of our proposed~\modelname~for PPCDR.


\section{Related Work}
\noindent\textbf{Deep Learning-based CDR.} CDR has been proposed to address the data sparsity problem in recommender systems by leveraging the denser information from related domains~\cite{zhu2021cross}.
Existing CDR models can not only model user-item relationships more accurately, but also bridge domain knowledge more effectively.
In the former, PPGN \cite{zhao2019cross} uses graph neural network to
model the high-order user-item interactions. 
In the latter, various techniques are utilized for bridging the information across domains,
such as cross connection~\cite{hu2018conet}, dual learning~\cite{li2020ddtcdr}, and domain adaption~\cite{liu2022collaborative}.
However, they overlook the importance of preserving the privacy of users, e.g., the interaction behaviors of users. 

\noindent\textbf{Privacy-preserving CDR.} 
User data in recommender systems are generally stored in two locations, i.e., (1) the \textit{end-user device} that generates data, and (2) the \textit{central database} that legally collects data.
The former first models data in decentralized devices, and then unifies the model parameters.
%
FCF~\cite{ammad2019federated} protects the behaviors of users by perturbing the local gradients.
MetaMF~\cite{lin2020meta} uses meta-learning to the federated recommender systems. 
%
But these solutions are inefficient in frequently communicating among different devices. 
The latter focuses on privately modeling data, and transferring the centralized data across domains.
UPC-SDG~\cite{Liu2022Privacy-Preserving} preserves privacy by synthesizing the less contributive items in single-domain recommendation. 
Several work focuses on transferring different kinds of knowledge, e.g., locations~\cite{gao2019privacy}
and social relationships~\cite{cui2021exploiting}, 
without leaking privacy.
Nevertheless, these work overlooks protecting the interaction data for recommendation in CDR. 
PriCDR is the first work to protect user rating information for transferring.
However, 
it causes looser privacy budget bound and unreliable recommendation performance.

\noindent\textbf{GAN-based Recommendation.} 
IRGAN \cite{wang2017irgan} is the first trial to apply GAN-based model to recommender systems, which generates user-item pairs and discriminates their relevance.  
CFGAN \cite{chae2018cfgan} proposes a model to generate the continuous purchase vectors, which avoids deteriorating the performance of discriminator caused by contradicting labels. 
Several work devotes to augmenting the rating data, e.g., (1) using the generated negative samples \cite{chae2019collaborative}, and (2) using auxiliary side information~\cite{wang2019enhancing}. 
GANMF \cite{dervishaj2022gan} notices that a single-valued feedback from discriminator is deficient in generating high-dimensional data, and models the difference between the real and fake profiles by an autoencoder like EBGAN~\cite{zhao2016energy}.
Most of these GAN-based models cannot be directly applied to \modelname, due to the several reasons, (1) overlooking privacy-preserving, (2) not stabilizing the modeling process, and (3) requiring raw data as input.

\section{Preliminaries of Differential Privacy}

\begin{definition}[\'Renyi Differential Privacy (RDP) \cite{mironov2017renyi}]
A mechanism $\mathcal{M}$ is $(\alpha, \varepsilon)-$RDP with order $\alpha \in (1, \infty)$ if for all neighboring datasets $\mathbf{x}$ and $\mathbf{x}^{\prime}$, the \'Renyi divergence satisfies: $D_{\alpha}\left(\mathcal{M}(\mathbf{x}) \| \mathcal{M}\left(\mathbf{x}^{\prime}\right)\right) \leq \varepsilon$, and further equals to $\left(\varepsilon+\frac{\log 1 / \delta}{\alpha-1}, \delta\right)-$DP, for any $\delta \in (0,1)$.  
\end{definition}

\begin{proposition}[RDP with Gaussian Mechanism \cite{dwork2014algorithmic,mironov2017renyi}]
\label{prop:gaussain_RDP}
For a $d$-dimensional function $f: \mathbf{x} \rightarrow  \mathbb{R}^{d}$ with sensitivity $\Delta_f= \max_{\mathbf{x}, \mathbf{x}^{\prime}} \|f(\mathbf{x}) - f(\mathbf{x}^{\prime})\|_{2}$, the Gaussian mechanism $\mathcal{M}_{\sigma_p, B} = f(\mathbf{x})+ \mathcal{N}(0,\sigma_p^2 B^2 \mathbf{I}) $, parameterized by the variance of Gaussian noise ${\sigma_p}$, is $(\alpha, \nicefrac{\alpha {\Delta_{f}}^2}{2 \sigma_p^2 B^2})-$RDP.
\end{proposition}

\section{Methodology}

\subsection{Framework of \modelname}
Firstly, we describe notations.
Without loss of generality, we assume there are two domains (i.e., $\mathcal{S}$ and $\mathcal{T}$) in CDR, which fully share the user set $\mathcal{U}$ with $N_\mathcal{U}$ users, but differ in rating matrices $\mathbf{R}_\mathcal{S}  \in \mathbb{R}^{N_\mathcal{U} \times N_{\mathcal{S}}}$ and $\mathbf{R}_\mathcal{T} \in \mathbb{R}^{N_\mathcal{U} \times N_{\mathcal{T}}}$ for $N_{\mathcal{S}}$ items in $\mathcal{S}$ and $N_{\mathcal{T}}$ items in $\mathcal{T}$, respectively.
Let $\boldsymbol{r}_i$, the $i$-th row of $\mathbf{R}$, be the user preference of user $i$. 
The goal of PPCDR is to transfer the information of $\mathbf{R}_\mathcal{S}$ in source domain to target domain \textbf{in a privacy-preserving way}, and further improve the recommendation performance with $\mathbf{R}_\mathcal{T}$. 

\begin{figure*}[t]
\centering
\includegraphics[width=2.1\columnwidth]{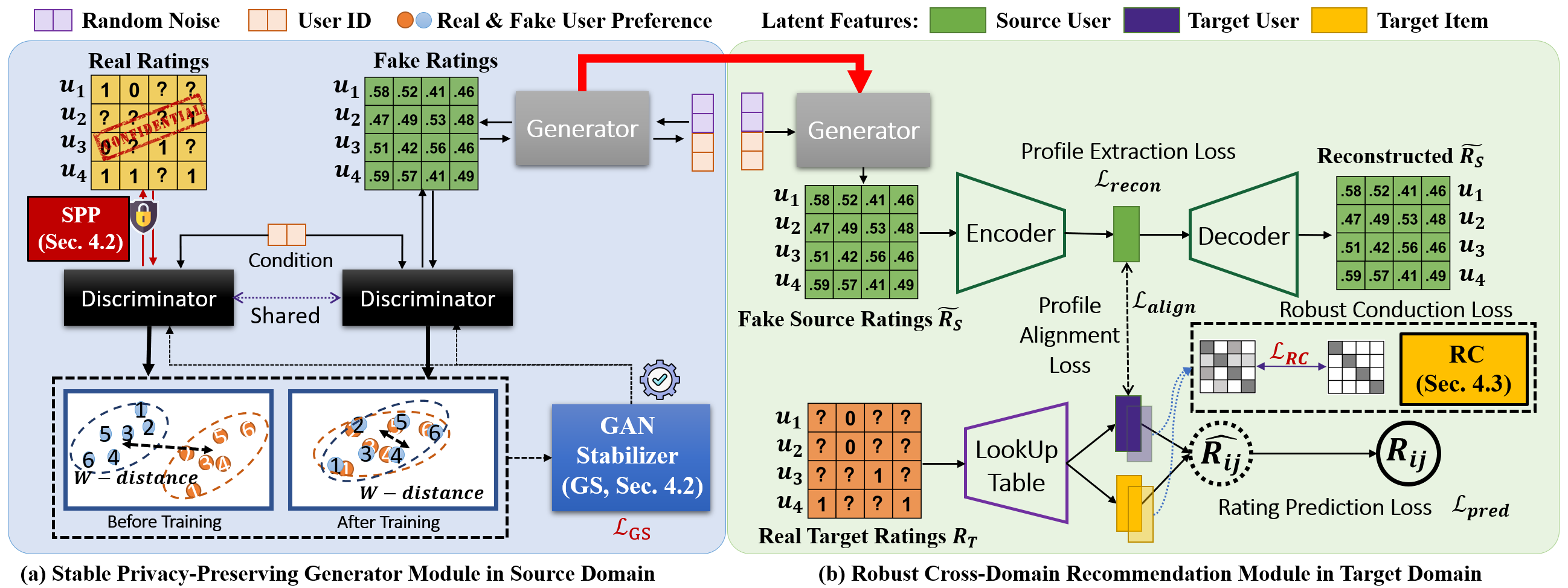} %
\caption{
Framework of \modelname.
%
}
\label{fig:framework}
\end{figure*}

To resolve PPCDR, we propose a framework named as \modelname~in this paper.
As Fig.~\ref{fig:framework} depicts, \modelname~includes two  modules, i.e., 
(1) \textit{stable privacy-preserving generator}~(\moduleAname) module in the source domain,
and (2) \textit{robust cross-domain recommendation}~(\moduleBname)  module in the target domain.
Specifically, \moduleAname~aims at stably modeling a privacy-preserving distribution of $\mathbf{R}_\mathcal{S}$ in source domain by selectively using RDP.
To leverage the information of user identifications ($\mathbf{u}_{id}$), we formulate \moduleAname~as a conditional GAN (CGAN) model and take $\mathbf{u}_{id}$ as conditions.
Then \moduleAname~publishes its \textit{generator} to target domain,
which is a flexible \textit{private data publisher} to generate perturbed $\mathbf{\widetilde{R}}_\mathcal{S}$ while allowing frequent queries in downstream tasks.
Thirdly, \moduleBname~robustly enhances the recommendation performance in target domain by leveraging  the perturbed knowledge $\mathbf{\widetilde{R}}_\mathcal{S}$ in source domain and the sparse information $\mathbf{R}_\mathcal{T}$ in target domain.

\subsection{Stable Privacy-Preserving Generator (SPPG)}
\nosection{Motivation}
In source domain, we aim to protect the private data by a \textit{private data publisher}, which isolates the source domain who provides data, and the target domain who requires data. 
Specifically, we train \moduleAname~using RDP, to obtain a privacy-preserving distribution of data, and publish the generator of \moduleAname~as a private data publisher for two benefits.
Firstly, the target domain gets access to the knowledge of source domain by sampling fake user preferences generated from the generator in \moduleAname, \textit{avoiding the risk of privacy exposure}. 
After well trained, the privacy budget will be a fixed value, 
since the generator in \moduleAname~does not need to visit the private data any more. 
Secondly, \moduleAname~\textit{augments the sparse rating data} in an adversarial way to better capture the user preferences in source domain. 

\nosection{Backbone of \moduleAname}
To start with, we introduce \moduleAname~in details.
As the left of Fig.~\ref{fig:framework} illustrates, \moduleAname~has a \textit{generator} ($\mathcal{G}$) and a \textit{discriminator} ($\mathcal{D}$), and takes user identifications $\mathbf{u}_{id}$ as conditions for both of them.
Specifically, $\mathcal{G}$~first concatenates a random noise vector $\boldsymbol{\sigma}_g \sim \mathcal{N}(0, \bm{I})$ and conditions $\mathbf{u}_{id}$.
Then $\mathcal{G}$ takes the concatenation as the input to construct fake user preferences $\tilde{\boldsymbol{r}}_\mathcal{S} = \mathcal{G}(\boldsymbol{\sigma}_g,\mathbf{u}_{i d})$. 
%
Given the conditions $\mathbf{u}_{id}$, $\mathcal{D}$ is encouraged to correctly classify the real user preference $\boldsymbol{r}_\mathcal{S}$ and the fake one $\tilde{\boldsymbol{r}}_\mathcal{S}$.
We measure the distribution divergence of $\boldsymbol{r}_\mathcal{S}$ and $\tilde{\boldsymbol{r}}_\mathcal{S}$ by Wasserstein distance
~\cite{arjovsky2017wasserstein}, which is good for modeling high-dimensional user preferences~\cite{dervishaj2022gan}.
To generate realistic user preferences, $\mathcal{G}$ cheats $\mathcal{D}$ by minimizing the Wasserstein distance.
On the contrary, $\mathcal{D}$ maximizes the Wasserstein distance to distinguish the distributions of fake user preferences and the real ones.
We maximize the loss of $\mathcal{L}_{\mathcal{G}}$ and $\mathcal{L}_{\mathcal{D}}$ iteratively to obtain the distribution of data in source domain:
\begin{equation}
\label{eq:dg}
\left\{
\begin{aligned}
&\mathcal{L}_\mathcal{G} = \mathbb{E}_{\tilde{\boldsymbol{r}} \sim p_{\tilde{\boldsymbol{r}}_\mathcal{S}}}\left[\mathcal{D}\left(\tilde{\boldsymbol{r}},\mathbf{u}_{i d}\right)\right],\\
&\mathcal{L}_\mathcal{D} = \mathbb{E}_{\boldsymbol{r} \sim p_{\boldsymbol{r}_\mathcal{S}}}\left[\mathcal{D}\left(\boldsymbol{r}, \mathbf{u}_{i d}\right)\right] - \mathbb{E}_{\tilde{\boldsymbol{r}} \sim p_{\tilde{\boldsymbol{r}}_\mathcal{S}}}\left[\mathcal{D}\left(\tilde{\boldsymbol{r}},\mathbf{u}_{i d}\right)\right].
\end{aligned}
\right.
\end{equation}

\nosection{Enhancing \moduleAname~with Privacy and Stability} 
Traditionally, we can directly perturb the gradients of \moduleAname~by adding random noise in the training procedure, e.g., adopting DP-SGD~\cite{abadi2016deep}, and get the distribution of private data in source domain.
However, it causes two challenges inevitably, i.e., \textit{(1) trade-off between privacy and performance}, and \textit{(2) stability of optimizing minmax objective}. 
The former means that adding too much noise will surely reduce the utility of the model, but may gain insignificant benefits of privacy-preserving.
The latter takes place because GAN-based models are deficient in stabilizing the process of modeling~\cite{roth2017stabilizing}, which will become worse when adding noise on the gradients for preserving privacy.
To tackle the two challenges, we devise selective privacy preserver (SPP) and GAN stabilizer (GS) for \moduleAname, which have a united optimization objective of selectively preserving privacy and stably modeling a GAN-based model.

\noindent\textit{Preserving privacy selectively for \moduleAname~with SPP.}
\label{module:spp}
To protect the private data in source domain and avoid severe performance degradation
in target domain, we devise SPP which uses RDP technique to selectively perturb the gradients of \moduleAname. 
As shown in the left of Fig.~\ref{fig:framework}, only $\mathcal{D}$ in \moduleAname~can directly access to the private real ratings.
Then generator obtains the knowledge of the private real ratings from the feedback of $\mathcal{D}$. 
Thus, SPP only needs to apply RDP on the gradients of $\mathcal{D}$, i.e., $\boldsymbol{g}_{\mathcal{D}}$, in back-propagation, and keeps the remaining gradients of \moduleAname~in private, according to the post-processing property of RDP~\cite{dwork2014algorithmic}. 

Additionally, we expand $\boldsymbol{g}_{\mathcal{D}}$ in \moduleAname~by chain rule: 
\begin{equation}
\label{eq:g_d}
\boldsymbol{g}_{\mathcal{D}}=\nabla_{\mathcal{D}} \mathcal{L}_{\mathcal{D}}\left(\boldsymbol{r}_\mathcal{S},\tilde{\boldsymbol{r}}_\mathcal{S};\boldsymbol{\theta}_{\mathcal{D}}, \mathbf{u}_{i d}\right) \cdot J_{\boldsymbol{\theta}_{\mathcal{D}}},
\end{equation}
where $J_{\boldsymbol{\theta}_{\mathcal{D}}}$ is the Jacobian matrix of $\mathcal{D}\left(\boldsymbol{r}_\mathcal{S},\tilde{\boldsymbol{r}}_\mathcal{S} ; \boldsymbol{\theta}_{\mathcal{D}},\mathbf{u}_{i d}\right)$. 
%
Hence we can simply perturb the gradients of the first layer in $\mathcal{D}$~\cite{chen2020gs} and preserve privacy.

Specifically, SPP implements RDP on $\boldsymbol{g}_{\mathcal{D}}$ to get its sanitized form  $\hat{\boldsymbol{g}}_\mathcal{D}$ by clipping and applying Gaussian mechanism on $\boldsymbol{g}_{\mathcal{D}}$.
Firstly, SPP bounds the sensitivity of optimization on training examples by clipping the gradient, i.e., $\nicefrac{\boldsymbol{g}}{\max (1, \nicefrac{ \|\boldsymbol{g}\|}{B} )}$ to ensure $ \|\boldsymbol{g}\| \leq B$, where $B$ is a clipping constant.
After that, SPP gets the sanitized form $\hat{\boldsymbol{g}}_\mathcal{D}$ by applying Gaussian mechanism
to perturb the values of $\nabla_{{\mathcal{D}}} \mathcal{L}_{\mathcal{D}}$:
 \begin{equation}
   \label{eq:sanitization}
\begin{aligned}
\hat{\boldsymbol{g}}_{\mathcal{D}}
&= \operatorname{SPP}(\nabla_{{\mathcal{D}}} \mathcal{L}_{\mathcal{D}}\left(\boldsymbol{r}_\mathcal{S},\tilde{\boldsymbol{r}}_\mathcal{S};\boldsymbol{\theta}_{\mathcal{D}},\mathbf{u}_{i d}\right)) \cdot J_{\boldsymbol{\theta}_{\mathcal{D}}}, \\
\end{aligned}
\end{equation}
where $\text{SPP}(\cdot) = \operatorname{clip}\left(\cdot \right)+\mathcal{N}\left(0, \sigma_{p}^{2} B^{2} \boldsymbol{I}\right) $, 
and $\sigma_{p}$ is the variance of noise in Gaussian mechanism. 
Therefore, SPP replaces $\boldsymbol{g}_{\mathcal{D}}$ (Eq.~\eqref{eq:g_d}) with $\hat{\boldsymbol{g}}_{\mathcal{D}}$ (Eq.~\eqref{eq:sanitization}) in updating, which preserves privacy of the modeling procedure selectively.

\begin{theorem}[Privacy Bound of Each \moduleAname ~Update Step]
\label{theorem:privacy_SPPG}
For batch size $N$, clipping constant $B$, and Gaussian mechanism $\mathcal{M}_{\sigma_{p}, B}$, each update step for training \moduleAname~satisfies
\begin{tiny}
$(\alpha, 2 N \alpha/\sigma_{p}^{2})-$RDP
\end{tiny} for $B=1$, and 
\begin{tiny}
$\left(2 N \alpha/\sigma_{p}^{2}+\frac{\log 1 / \delta}{\alpha-1}, \delta\right)-$DP. 
\end{tiny}
\begin{proof}

The RDP of updating a single user is $(\alpha, 2\alpha/\sigma_{p}^{2})-$RDP with $B=1$ in Gaussian mechanism, i.e., $\mathcal{M}_{\sigma_{p}, 1} = \text{clip}(\boldsymbol{g}_{\mathcal{D}}) + \mathcal{N}\left(0, \sigma_{p}^{2} \boldsymbol{I}\right)$. 
Then, the privacy amplification of updating a batch of users could be obtained by using the composition of $N$ Gaussian mechanisms~\cite{dwork2008differential}.
Thus, we obtain the RDP privacy upper bound for a batch of users, i.e.,
$(\alpha, 2 N \alpha/\sigma_{p}^{2})-$RDP, which is measured by $\alpha-$order \'Renyi divergence as bellow:
\begin{small}
\begin{equation}
\begin{aligned}
D_{\alpha}\left(\hat{\boldsymbol{g}}_{\mathcal{D}}(\boldsymbol{r}_{1:N}), \hat{\boldsymbol{g}}_{\mathcal{D}}\left(\boldsymbol{r}_{1:N}^{\prime}\right)\right) \leq N \cdot 2 \alpha / \sigma_{p}^{2}.
\end{aligned}
\end{equation}
\end{small}
According to \cite{dwork2014algorithmic} and \cite{mironov2017renyi}, the DP privacy upper bound further equals to $\left(2 N \alpha/\sigma_{p}^{2}+\frac{\log 1 / \delta}{\alpha-1}, \delta\right)-$DP.
\end{proof} 
\end{theorem}

\noindent\textit{Stabilizing the modeling process of \moduleAname~with GS.}
\label{module:gs}
Note that SPP applies Gaussian mechanism with noise $\sigma_{\text{gr}} $, ranging from $[-\sigma_{p} B, \sigma_{p} B]$, on the gradients in \moduleAname.
In this way, the ground truth distribution of user preferences is not definite during the modeling procedure. 
To simulate the perturbed ground truth distribution, we first simplify \moduleAname~to a Dirac GAN~\cite{mescheder2018training}, which is widely studied for analyzing the stability of GAN-based model.
In Dirac GAN, the ground truth distribution of $\boldsymbol{r}$ is $p(\boldsymbol{r})=\text{Dirac}(\boldsymbol{r}-\boldsymbol{c})$, where all elements of the vector $\boldsymbol{c}$ are constant. 
%
Then we substitute the vector $\boldsymbol{c}$ in $p(\boldsymbol{r})$ with $\boldsymbol{m}(t) = \boldsymbol{c} -(1-2 \phi)\boldsymbol{\sigma}_{\text{gr}}$ $(\phi \in [0,1])$
by interpolating the noise brought from SPP. 
In terms of Eq.~\eqref{eq:dg}, \moduleAname~aims to estimate the distribution of perturbed ground truth $p(\boldsymbol{r})=\text{Dirac}(\boldsymbol{r}-\boldsymbol{m}(t))$ as follows:
\begin{small}
\begin{equation}
\label{eq:opt_dp_dg}
\frac{d \boldsymbol{\theta}_{\mathcal{D}}}{d t} =\boldsymbol{m}(t)-\boldsymbol{\theta}_{\mathcal{G}}(t),  \quad \quad \frac{d \boldsymbol{\theta}_{\mathcal{G}}}{d t} =\boldsymbol{\theta}_{\mathcal{D}}(t).
\end{equation}
\end{small}

From the perspective of control theory, we can treat Eq.~\eqref{eq:opt_dp_dg} as a dynamic system, which takes $\boldsymbol{m}(t)$ as input, and $\boldsymbol{y}(t)=(\boldsymbol{\theta}_\mathcal{D}( t),\boldsymbol{\theta}_\mathcal{G}(t))$, $\forall t>0$ as output.
And we figure out that the modeling procedure of \moduleAname~in Eq.~\eqref{eq:opt_dp_dg} is not a stable dynamic system. 
Moreover, the stability of modeling $\mathcal{G}$ depends on the stability of modeling $\mathcal{D}$. 

Motivated by~\cite{xu2020understanding}, we take GS, as a controller parameterized by $\tau~(\tau > 0)$, 
to formulate a closed-loop control system~\cite{kailath1980linear}, which is widely used to improve the stability of a non-linear dynamic system like Eq.~\eqref{eq:opt_dp_dg}: 
\begin{equation}
\label{eq:stable_d}
\frac{d \boldsymbol{\theta}_{\mathcal{D}}}{d t} = \boldsymbol{m}(t) -\boldsymbol{\theta}_\mathcal{G} (t) - \tau \boldsymbol{\theta}_\mathcal{D} (t).
\end{equation}

Comparing Eq.~\eqref{eq:opt_dp_dg} and Eq.~\eqref{eq:stable_d}, we find that GS brings an extra optimization objective to the loss function of $\mathcal{D}$: 
\begin{small}
\begin{equation}\nonumber
\begin{aligned}
\mathcal{L}_{\mathrm{GS}}= - \frac{\tau}{2}\left(\mathbb{E}_{\boldsymbol{r} \sim  p_{\boldsymbol{r}_\mathcal{S}} }\left[\mathcal{D}^{2}\left(\boldsymbol{r}, \mathbf{u}_{i d}\right)\right]+\mathbb{E}_{{\tilde{\boldsymbol{r}} \sim p_{\tilde{\boldsymbol{r}}_\mathcal{S}}} }\left[\mathcal{D}^{2}\left( \tilde{\boldsymbol{r}}, \mathbf{u}_{i d}\right)\right]\right).
\end{aligned}
\end{equation}
\end{small}

Intuitively, we can find that the loss of GS $\mathcal{L}_{\text{GS}}$ is actually a regularization term, which penalizes the abrupt outputs from $\mathcal{D}$.
Given the feedback of $\mathcal{L}_{\text{GS}}$, \moduleAname~adjusts its $\mathcal{D}$ to measure the distance of the real and fake user preferences ($\boldsymbol{r}$ and $\tilde{\boldsymbol{r}}$) in a tighter latent representation space,
which stably estimates the distribution of private data in source domain.

\nosection{Optimization in \moduleAname}
Finally, derived from the Eq.~\eqref{eq:dg}, we can achieve a stable modeling procedure for \moduleAname~by maximizing the loss of \moduleAname~with SPP and GS: 
\begin{small}
\begin{equation} 
\mathbf{L}_\mathcal{G} = \mathcal{L}_\mathcal{G},\quad
\mathbf{L}_\mathcal{D} = \mathcal{L}_\mathcal{D} +  \mathcal{L}_{\text{GS}}.
\end{equation}
\end{small}

In summary, SPP (red block in Fig.~\ref{fig:framework}) preserves the privacy of data in source domain without decreasing the performance severely, and GS (blue block in Fig.~\ref{fig:framework}) stabilizes the modeling procedure of \modelname~to better capture the distribution of private data in source domain.

\subsection{Robust CDR (RCDR)}
\nosection{Motivation}
After privately modeling the distribution of data by \moduleAname~in source domain, we introduce the way of leveraging it in target domain.
Specifically, we devote to leverage the knowledge contained in the generator published from \moduleAname, and improve the recommendation performance with the raw data of target domain.
\moduleBname~should extract user preferences from the published generator and combine the user preferences of two domains together to enhance the target domain recommendation. 

\nosection{Backbone of \moduleBname}
We introduce the general overview of \moduleBname~in the right of Fig.~\ref{fig:framework}, which contains (1) a \textit{generator} published from \moduleAname, and (2) a \textit{CDR module} to leverage the perturbed information from source domain and perform recommendation in target domain.
Additionally, CDR module consists of three sub-modules, i.e., a \textit{profile extraction} module, a \textit{recommendation prediction} module, and a \textit{profile alignment} module.
Firstly, given user identification $\mathbf{u}_{id}$ and a random noise vector $\boldsymbol{\sigma}_g \sim \mathcal{N}(0, \bm{I})$, the well-trained generator samples perturbed, but effective user preferences  $\tilde{\boldsymbol{r}}_\mathcal{S} = \mathcal{G}(\boldsymbol{\sigma}_g,\mathbf{u}_{i d})$.
Then the \textit{profile extraction} module reconstructs $\tilde{\boldsymbol{r}}_{\mathcal{S}}$ to extract latent feature as user profile  ${\boldsymbol{z}_\mathcal{S}} = \operatorname{Enc} (\tilde{\boldsymbol{r}}_{\mathcal{S}})$ (${\boldsymbol{z}_\mathcal{S}} \in \mathbb{R}^{1 \times K}$ with $K$ denoting the dimension size of the source domain), which is empirically useful for target model~\cite{chen2022differential}.
To capture a batch of user profiles in source domain $\boldsymbol{Z}_{\mathcal{S}} = \left[ \boldsymbol{z}_{\mathcal{S}, 1}, \ldots , \boldsymbol{z}_{\mathcal{S}, N} \right] \in \mathbb{R}^{N \times K}$, we minimize the reconstruction loss as follows:
\begin{equation}
    \label{eq:recon} 
    \mathcal{L}_{\text{recon}} = \sum_{u=1}^N \left\| \operatorname{Dec}\left(\operatorname{Enc}\left (\tilde{\boldsymbol{r}}_{\mathcal{S},u}\right)\right) -\tilde{\boldsymbol{r}}_{\mathcal{S},u} \right\|^2_F,
\end{equation} 
where $N$ is the batch size.
Meanwhile, a \textit{recommendation prediction} module, i.e., DMF~\cite{xue2017deep}, learns a lookup table from $\mathbf{R}_\mathcal{T}$ in the target domain, 
to obtain a batch of user profiles $\boldsymbol{Z}_{\mathcal{T}} = \left[ \boldsymbol{z}_{\mathcal{T}, 1}, \ldots , \boldsymbol{z}_{\mathcal{T}, N} \right]$ and the item features
$\boldsymbol{V}_{\mathcal{T}} = \left[ \boldsymbol{v}_{\mathcal{T}, 1} \ldots , \boldsymbol{v}_{\mathcal{T}, N} \right]$, where  ${\boldsymbol{Z}_\mathcal{T}}, {\boldsymbol{V}_\mathcal{T}} \in \mathbb{R}^{N \times K}$.
Additionally, we bridge the knowledge of source domain to the target domain by \textit{profile alignment} module, which minimizes the user dissimilarities in both domains (${\boldsymbol{z}_\mathcal{S}}$ and ${\boldsymbol{z}_\mathcal{T}}$): 
\begin{equation}
    \label{eq:align_preference} 
\mathcal{L}_{\text{align}} = \sum_{u=1}^N \left\|  {\boldsymbol{z}_{\mathcal{S}, u}} - {\boldsymbol{z}_{\mathcal{T},u}} \right\|^2_F.
\end{equation}
Lastly, \textit{recommendation prediction} module predicts user preference based on user profile $\boldsymbol{z}_{\mathcal{T}}$ and item property ${\boldsymbol{v}_{\mathcal{T}}}$ 
by minimizing binary cross entropy $\operatorname{F}_{\text{bce}}(\cdot)$ as below: 
\begin{equation}
    \label{eq:pred} 
    \mathcal{L}_{\text{pred}} =  \sum_{u=1}^N \operatorname{F}_{\text{bce}} ( \widehat{\boldsymbol{r}}_{\mathcal{T}, {u}},  \boldsymbol{r}_{\mathcal{T}, u}), 
\end{equation}
where $\widehat{\boldsymbol{r}}_{\mathcal{T}, {u}} = {\boldsymbol{z}_{\mathcal{T},u}}^{\top} {\boldsymbol{v}_{\mathcal{T}}}$ is the predicted user preference of user $u$ in target domain.

\nosection{Enhancing \moduleBname~with Robustness} 
Few existing work in CDR prevents the performance from deteriorating in target domain when utilizing the information from other domains~\cite{gao2022self}.
In PPCDR, source domain inevitably transfers the perturbed data to target domain for the sake of preserving privacy, which potentially decreases the performance of models in target domain.

\noindent\textit{Robustifying the recommendation performance in \moduleBname~with RC.}
\label{module:rc}
We propose a robustness conductor (RC) in \moduleBname~to robustly enhance the performance of recommendation model in target domain.
Specifically, RC minimizes the impact of noisy data generated by the published generator from \moduleAname, which empirically causes the latent features, i.e., $\boldsymbol{Z}_{\mathcal{S}}$, $\boldsymbol{Z}_{\mathcal{T}}$, and $\boldsymbol{V}_{\mathcal{T}}$,  redundant and dependent in a batch of samples.

To enhance robustness, RC disentangles the features of different dimensions in $\boldsymbol{Z}_{\mathcal{T}}$ and $\boldsymbol{V}_{\mathcal{T}}$, which are directly used for predicting user preferences in target domain. 
The user features $\boldsymbol{Z}_{\mathcal{T}}$ and the item features $\boldsymbol{V}_{\mathcal{T}}$ have the same processing procedure. 
Take user features as an example, RC first uses Z-score normalization~\cite{lv2022causality} to every column (dimension) of the batched user features $\boldsymbol{Z}_{\mathcal{T}}$ to get $\bar{\boldsymbol{Z}}_{\mathcal{T}}$, whose $i$-th column (dimension) is denoted as ${\bar{\boldsymbol{z}}_{\mathcal{T}, i}}$ ($i \leq K$). 
Then RC measures the cross correlation matrix among different columns of 
normalized user features, i.e., $\boldsymbol{C}_{\bar{\boldsymbol{Z}}_{\mathcal{T}}} \in \mathbb{R}^{K \times K}$, where  $\boldsymbol{C}_{\bar{\boldsymbol{Z}}_{{\mathcal{T}}, {i j}}}=\nicefrac{\left\langle{\bar{\boldsymbol{z}}_{\mathcal{T},i}}, {\bar{\boldsymbol{z}}_{\mathcal{T},j}}\right\rangle}{\left\|{\bar{\boldsymbol{z}}_{\mathcal{T},i}}\right\|\left\|{\bar{\boldsymbol{z}}_{\mathcal{T},j}}\right\|}$  (for $i,j \in 1,\ldots, K$). 
Intuitively, RC disentangles the redundant representations in a batch from two aspects: (1) keeping the representations of the same dimension consistent, and (2) enforcing the representations of different dimensions mutually independent.
To realize the former, RC approximates the diagonal elements of cross-correlation matrix $\boldsymbol{C}_{\bar{\boldsymbol{Z}}_{{\mathcal{T}}, {i i}}}$ to $1$. 
To reach the latter, RC approximates  
$\boldsymbol{C}_{\bar{\boldsymbol{Z}}_{{\mathcal{T}}, {i j}}}$ (for $i \neq j$) to $0$. 
Similarly, RC processes the latent \textit{items} representations $\boldsymbol{V}_{\mathcal{T}}$ to get its cross correlation matrix, i.e.,
$\boldsymbol{C}_{\bar{\boldsymbol{V}}_{{\mathcal{T}}}} \in \mathbb{R}^{K \times K}$,
and disentangles the features of different dimensions. 
Finally, we have robust regularization for \moduleBname~to minimize:
\begin{equation}
\label{eq:cl}
\begin{aligned}
    \mathcal{L}_{RC} 
    & =  \|\boldsymbol{C}_{\bar{\boldsymbol{Z}}_{\mathcal{T}}} -\boldsymbol{I}\|_{F}^{2} + \|\boldsymbol{C}_{\bar{\boldsymbol{V}}_{\mathcal{T}}}-\boldsymbol{I}\|_{F}^{2},
\end{aligned}
\end{equation}
where 
$\boldsymbol{I} \in \mathbb{R}^{K \times K}$ is the identity matrix.
Therefore, RC can make the noisy and redundant representations into clean and independent ones, and further enhance recommendation prediction in a robust way.
RC is marked in the right of Fig.~\ref{fig:framework} as a yellow block, which is a flexible plugin to improve the robustness of recommendation models.

\nosection{Optimization in \moduleBname}
Overall, the optimization of \moduleBname~is to minimize:
\begin{equation}
\label{opt:cdrpg}
\begin{aligned}
\mathbf{L}_{\text{\moduleBname}} &= \mathcal{L}_{\text{recon}} + \mathcal{L}_{\text{pred}} +  \lambda_{A}\mathcal{L}_{\text{align}} + \lambda_{R} \mathcal{L}_{\text{RC}},
\end{aligned}
\end{equation}
where $\lambda_{A}$ and $\lambda_{R}$ are hyper-parameters to balance different types of losses.
\modelname~can stably model the distribution of private data in source domain and robustly enhance the recommendation performance in the target domain. 

\section{Experiments and Discussion}

\begin{table*}[h] 
\renewcommand\arraystretch{0.5}
\centering
\Huge

\resizebox{\textwidth}{!}{
\begin{tabular}{ccccccccccccc}
\toprule
            \multirow{1}{*}{}  & \multicolumn{6}{c}{\textbf{Douban} book $\rightarrow$ music}             & \multicolumn{6}{c}{\textbf{Amazon} music  $\rightarrow$  book}            \\
              \midrule
             & \textbf{HR@5}   & \textbf{NDCG@5} & \textbf{MRR@5}  & \textbf{HR@10}   & \textbf{NDCG@10} & \textbf{MRR@10} &   \textbf{HR@5}   & \textbf{NDCG@5} & \textbf{MRR@5}  & \textbf{HR@10}   & \textbf{NDCG@10} & \textbf{MRR@10}  \\  
 \midrule
BPR   & .1284 & .0984 & .0556 & .2364 & .1221 & .0917 & .2556 & .1981 & .1887 & .3711 & .2688 & .2473 \\
NeuMF & .1317 & .0916 & .0677 & .2416 & .1375 & .1032 & .2660 & .2177 & .1952 & .4023 & .2729 & .2536 \\
DMF   & .1678 & .1143 & .1029 & .2533 & .1377 & .1014 & .2735 & .2435 & .2171 & .4216 & .2805 & .2703 \\
CFGAN & .1234 & .0745 & .0723 & .2024 & .0999 & .0769 & .1168 & .0752 & .0617 & .2071 & .1009 & .0722 \\
GANMF & .1385 & .0879 & .0714 & .2219 & .1146 & .0824 & .0632 & .0369 & .0285 & .1180 & .0546 & .0357 \\
\hline
CoNet  & .2042 & .1325 & .1082 & .3128 & .1611 & .1282 & .3616 & .2824 & .2564 & .4693 & .3172 & .2707 \\
DDTCDR & .2100 & .1446 & .1116 & .3315 & .1701 & .1329 & .4091 & .3199 & .2823 & .4864 & .3540 & .3296 \\

ETL          & .2781 & .1974 & .1691 & .4048 & .2379 & .1857 & .4883 & .3798 & .3439 & .6092 & .4188 & .3600 \\
DARec        & .2056 & .1346 & .1108 & .3149 & .1677 & .1243 & .2975 & .2119 & .1846 & .4175 & .2503 & .1995 \\


PriCDR & \underline{.3074} & \underline{.2107} & \underline{.1791} & \underline{.4448} & \underline{.2549} & \underline{.1972} & \underline{.5605} & \underline{.4527} & \underline{.4177} & \underline{.6663} & \underline{.4875} & \underline{.4321} \\
 \midrule
\modelname-\textbf{GS} & .2944 & .1958 & .1693 & .4448 & .2434 & .1885 & .5701 & .4639 & .4289 & .6767 & .4986 & .4432 \\
\modelname-\textbf{RC} & .3084 & .2123 & .1808 & .4686 & .2639 & .2022 & .2862 & .1953 & .1659 & .4175 & .2375 & .1832 \\
\modelname & \textbf{.3701} & \textbf{.2691} & \textbf{.2362 } & \textbf{.5097 } & \textbf{.3116 } & \textbf{.2532 } & \textbf{.6012 } & \textbf{.4886 } & \textbf{.4514} & \textbf{.7078} & \textbf{.5231} & \textbf{.4657} \\  
\textbf{Improvement} ($\uparrow$) & \textbf{ 20.40\%} & \textbf{27.72\%} & \textbf{31.88\%} & \textbf{14.59\%} & \textbf{ 22.24\%} & \textbf{28.40\%}  & \textbf{7.26\%} & \textbf{7.93\%} & \textbf{8.07\%} & \textbf{6.23\%} & \textbf{7.30\%} & \textbf{7.78\%} \\
\midrule
           \multirow{1}{*}{}  & \multicolumn{6}{c}{\textbf{Douban} movie  $\rightarrow$  book}          &  \multicolumn{6}{c}{\textbf{Amazon} movie  $\rightarrow$ book}              \\  \midrule
             
BPR   & .1835 & .1100 & .1041 & .2858 & .1440 & .1285 & .2642 & .1981 & .1887 & .3837 & .2169 & .2247 \\
NeuMF & .1902 & .1266 & .1093 & .3164 & .1562 & .1302 & .2866 & .1922 & .1920 & .3923 & .2443 & .2581 \\
DMF   & .2132 & .1220 & .1170 & .3536 & .1625 & .1488 & .3035 & .2144 & .2417 & .4116 & .2568 & .2670  \\
CFGAN & .1760 & .1113 & .0873 & .2745 & .1363 & .0977 & .1377 & .0885 & .0762 & .2176 & .1143 & .0839 \\
GANMF & .1131 & .0841 & .0746 & .1557 & .0978 & .0802 & .1169 & .0779 & .0653 & .1883 & .1010 & .0747 \\
\hline
CoNet  & .2516 & .1681 & .1517 & .3432 & .2316 & .1782 & .3611 & .2571 & .3095 & .4366 & .3088 & .3242 \\
DDTCDR & .2638 & .1793 & .1603 & .3699 & .2481 & .1837 & .4393 & .3434 & .3231 & .4838 & .3476 & .3427 \\

ETL          & .3571 & .2432 & .2091 & .4981 & .2922 & .2294 & .5436 & .4288 & .3909 & .6626 & .4673 & .4068 \\
DARec         & .2700 & .1871 & .1611 & .3958 & .2286 & .1771 & .3665 & .2736 & .2430 & .4950 & .3150 & .2600 \\
PriCDR & \underline{.3577} & \underline{.2434} & \underline{.2101} & \underline{.4987} & \underline{.2922} & \underline{.2301} & \underline{.5862} & \underline{.4868} & \underline{.4538} & \underline{.6889} & \underline{.5200} & \underline{.4675} \\
 \midrule
\modelname-\textbf{GS} & .3647 & .2640 & .2323 & .5114 & .3088 & .2510 & .5976 & .4970 & .4633 & .6923 & .5275 & .4759 \\
\modelname-\textbf{RC} & .4123 & .2976 & .2572 & .5654 & .3457 & .2768 & .3614 & 0.2641 & .2322 & .5003 & .3088 & .2505  \\
\modelname  &  \textbf{.4238} & \textbf{.3050} & \textbf{.2655} & \textbf{.5680} & \textbf{.3524} & \textbf{.2849} & \textbf{.6253} & \textbf{.5218} & \textbf{.4874} & \textbf{.7181} & \textbf{.5518} & \textbf{.4999}  \\  
\textbf{Improvement} ($\uparrow$) & \textbf{ 18.48\%} & \textbf{25.31\%} & \textbf{26.37\%} & \textbf{13.90\%} & \textbf{20.60\%} & \textbf{23.82\%}  & \textbf{6.67\%} & \textbf{7.19\%} & \textbf{7.40\%} & \textbf{4.24\%} & \textbf{6.12\%} & \textbf{6.93\%} \\
\midrule
            \multirow{1}{*}{} & \multicolumn{6}{c}{\textbf{Douban} movie  $\rightarrow$ music}           & \multicolumn{6}{c}{\textbf{Amazon} movie $\rightarrow$ music}             \\ \midrule
BPR   & .1535 & .0920 & .0941 & .2958 & .1640 & .1115 & .2443 & .1578 & .1631 & .3793 & .2238 & .1971 \\
NeuMF & .1625 & .1005 & .1061 & .3027 & .1498 & .1220 & .2664 & .2137 & .1801 & .4052 & .2372 & .2188 \\
DMF   & .1872 & .1023 & .1039 & .3060 & .1712 & .1267 & .3427 & .2219 & .1958 & .4126 & .2528 & .2375  \\
CFGAN & .1422 & .1850 & .1845 & .2351 & .1983 & .1895 & .0966 & .0618 & .0484 & .1694 & .0833 & .0571 \\
GANMF & .1213 & .0761 & .0626 & .2114 & .0966 & .0710 & .0795 & .0463 & .0356 & .1444 & .0672 & .0441  \\
\hline
CoNet  & .2331 & .1347 & .1322 & .3217 & .1924 & .1355 & .3712 & .2824 & .2964 & .4693 & .3172 & .2707 \\
DDTCDR & .2456 & .1572 & .1518 & .3877 & .2163 & .1421 & .3819 & .3508 & .3148 & .4877 & .3860 & .3389  \\

ETL       & .3270 & .2243 & .1908 & \underline{.4578} & .2665 & .2082 & .5561 & .4300 & .3884 & .6792 & .4706 & .4052 \\
DARec         & .2427 & .1626 & .1353 & .3469 & .1949 & .1505 & .3890 & .3859 & .3519 & .5170 & .4280 & .3693 \\
PriCDR & \underline{.3346} & \underline{.2305} & \underline{.1981} & .4398 & \underline{.2689} & \underline{.2138} & \underline{.6015} & \underline{.4872} & \underline{.4498} & \underline{.7079} & \underline{.5216} & \underline{.4640}  \\
\midrule
\modelname-\textbf{GS} & .3318 & .2268 & .1972 & .4777 & .2740 & .2171 & .6071 & .4905 & .4515 & .7200 & .5267 & .4662 \\
\modelname-\textbf{RC} & .2739 & .1911 & .1658 & .4095 & .2345 & .1836 & .3164 & .2159 & .1828 & .4536 & .2602 & .2010 \\
\modelname     & \textbf{.3782} & \textbf{.2727} & \textbf{.2335} & \textbf{.5185} & \textbf{.3181} & \textbf{.2512} & \textbf{.6379} & \textbf{.5187} & \textbf{.4779} & \textbf{.7460} & \textbf{.5529} & \textbf{.4921} \\
\textbf{Improvement} ($\uparrow$) & \textbf{ 13.03\%} & \textbf{18.31\%} & \textbf{17.87\%} & \textbf{13.26\%} & \textbf{18.30\%} & \textbf{17.49\%}  & \textbf{6.05\%} & \textbf{6.47\%} & \textbf{6.25\%} & \textbf{5.38\%} & \textbf{6.00\%} & \textbf{6.06\%} \\
 \bottomrule
\end{tabular}
}
\caption{Experimental results of target domain on Douban and Amazon datasets. We bold  the best result, underline the runner-up, and all of the improvements are significant with \textit{p-value} $< 0.01$ using the paired sample t-test. 
} 
\label{tb:exp_res}
\end{table*}

In this section, we aim to answer the following questions through empirical studies:
\textbf{Q1}: Can \modelname~outperform existing single-domain recommendation models, the state-of-the-art (SOTA) CDR models in plaintext, and the SOTA PPCDR models? \textbf{Q2}: How can GS and RC contribute to the \textit{performance} of \modelname? \textbf{Q3}: How can SPP preserve privacy in \modelname~\textit{in a cost-effective way}? \textbf{Q4}: How do hyper-parameters impact \modelname? 

\subsection{Experimental Setup}
\textbf{Datasets.} We use two datasets, i.e., \textbf{Amazon}
\cite{ni2019justifying} and \textbf{Douban}
\cite{zhu2021unified}.
Following \cite{chen2022differential}, we first select three domains on Amazon dataset, i.e., Movies and TV (Movie), Books (Book), and CD Vinyl (Music), and three domains on Douban dataset, i.e., Book, Music, and Movie. 
We binarize the ratings larger or equal to 3 as positive, and others as negative. 
We filter the users and items with less than 5 interactions. 

\noindent\textbf{Comparison methods.} 
%
We compare \modelname~with  the following representative models: 
(1) \textbf{BPR} \cite{rendle2009bpr} optimizes matrix factorization with implicit feedback by a pairwise ranking loss. 
(2) \textbf{NeuMF} \cite{he2017neural} is a neural collaborative filtering method which uses multilayer perceptron (MLP) to learn the inner product for prediction.
(3) \textbf{DMF} \cite{xue2017deep} is a deep matrix factorization model that maps both users and items into a low-dimensional space with non-linear projection.
(4) \textbf{CFGAN} \cite{chae2018cfgan} generates the user purchase vector based on conditional GAN.
(5) \textbf{GANMF} enhances the feedback of discriminator derived from energy-based GAN~\cite{zhao2016energy}.
(6) \textbf{CoNet} \cite{hu2018conet} introduces a cross connection unit to enable dual knowledge transfer across domains. 
(7) \textbf{DDTCDR} \cite{li2020ddtcdr} introduces deep dual learning to transfer knowledge in CDR.
(8) \textbf{DARec} \cite{yuan2019darec} shares user representation across different domains via domain adaption.
(9) \textbf{ETL} \cite{chen2020towards} adopts the equivalent transformation to selectively share domain-related and domain-specific knowledge.
(10) \textbf{PriCDR} \cite{chen2022differential} is the SOTA PPCDR model that publishes the perturbed interaction data using DP to target domain.

\noindent\textbf{Implementation details.} 
We tune the hyper-parameters of \modelname~and baseline models to their best values for a fair comparison. 
Specifically, we choose RMSprop~\cite{tieleman2012lecture} as the optimizer for \moduleAname, and Adam~\cite{kingma2014adam} for \moduleBname. 
We set batch size $N$ as 128, learning rate $\eta =0.01$ for Douban, and $\eta =0.0005$ for Amazon.
We set clipping constant $B = 1$, and the dimension of latent features $K=200$. 
We evaluate the top 5 and 10 of ranking results by Hit Ratio~(HR), Normalized Discounted Cumulative Gain (NDCG), and Mean Reciprocal Rank (MRR).

\begin{figure}[h]
\subfigure[\textbf{Douban} movie $\rightarrow $ book]{
    \begin{minipage}{0.46\columnwidth}{
\centering
\label{fig:dp_douban_m2mu}
\includegraphics[scale=0.5]{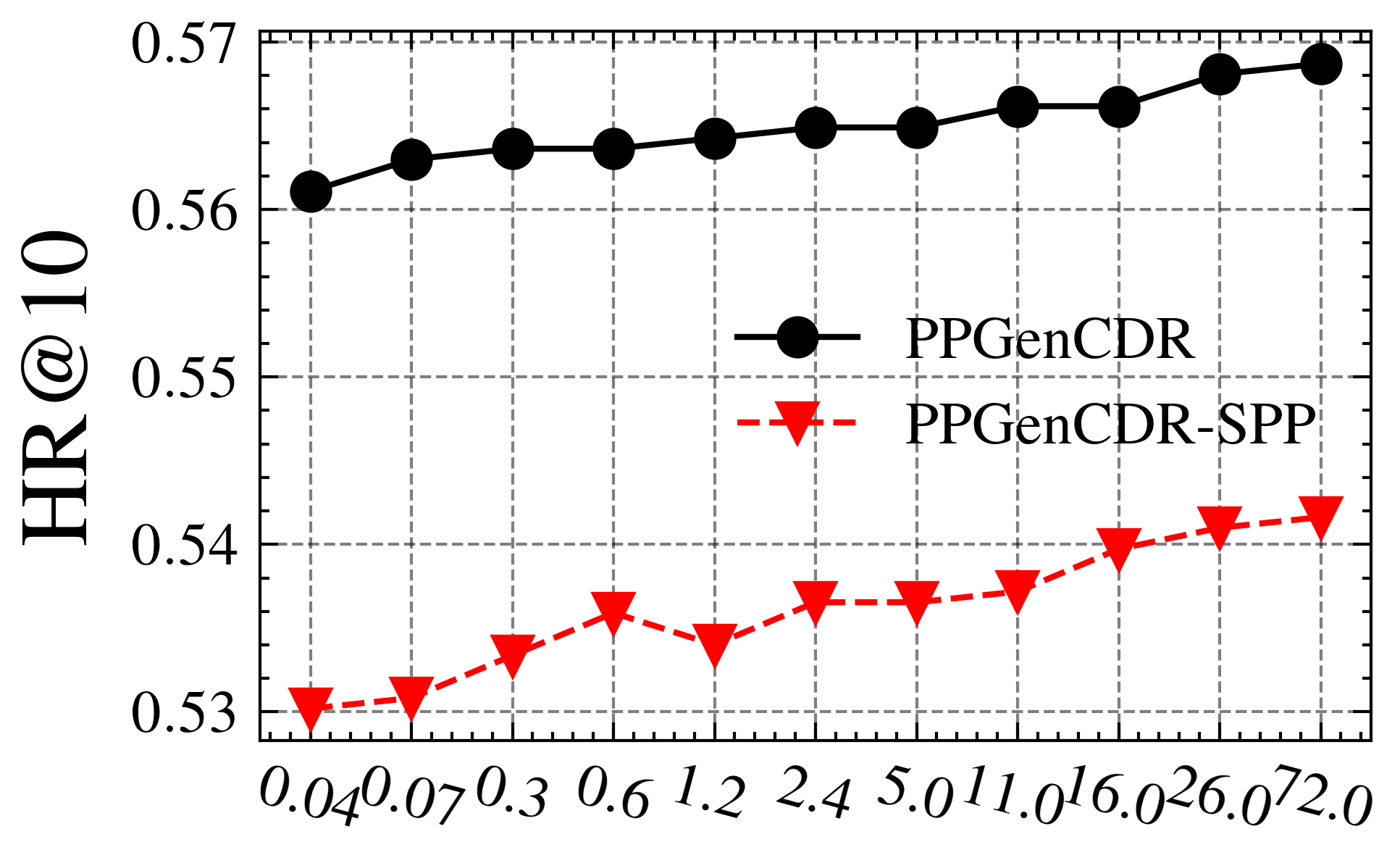} %
}
\end{minipage}
}
\subfigure[\textbf{Amazon} movie $\rightarrow $ music]{
\begin{minipage}{0.46\columnwidth}{
\centering
\label{fig:dp_amazon_m2mu}
\includegraphics[scale=0.5]{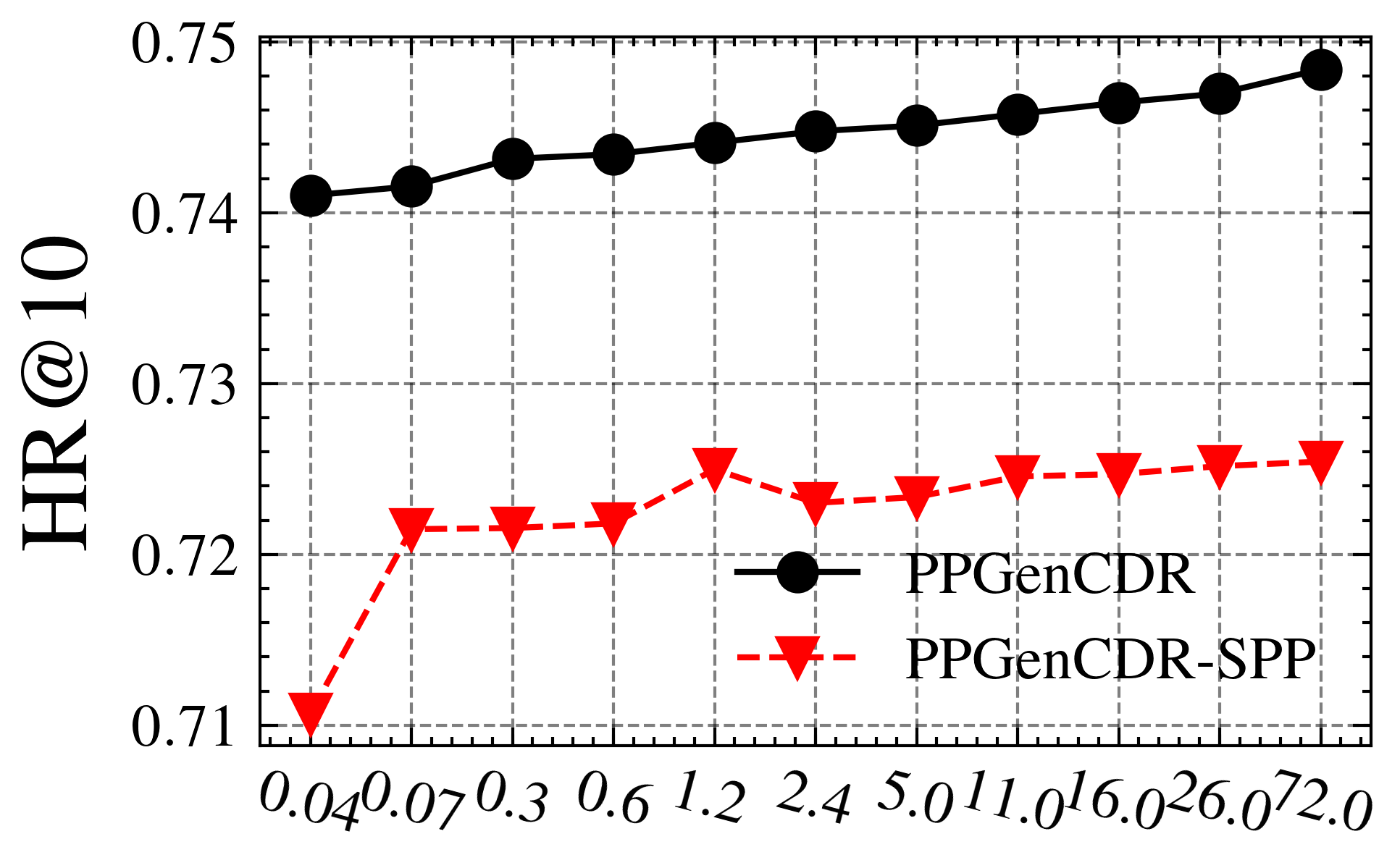} %
}
\end{minipage}
}
\caption{Privacy analysis of $\varepsilon$ on \modelname. }
\label{fig:privacy}
\end{figure}

\subsection{Recommendation Performance Comparison}
\nosection{Model Comparison (Q1)}
We report the comparison results on \textbf{Douban} and \textbf{Amazon} datasets in the Tab. \ref{tb:exp_res}, where $A \rightarrow B$ means transferring knowledge of domain $A$ to domain $B$.
%
%
We can find that: (1) Compared with the single-domain models, all of the CDR models contribute to alleviating the sparsity issue by transferring the knowledge from other domains.
(2) Compared with the SOTA CDR models in plaintext, \modelname~and PriCDR also achieve 
comparable performance, indicating that it is possible to find a trade-off between preserving privacy and maintaining performance.
(3) \modelname~significantly outperforms the SOTA PPCDR model (i.e., PriCDR) by at least 13.03\% and 6.05\% on Douban and Amazon in terms of HR@5, respectively, illustrating that \modelname~breaks the performance bottleneck by enhancing model stability and robustness.

\begin{figure}[t]
\subfigure[Effect of regularizing $\lambda_{A}$]{
\begin{minipage}{0.47\columnwidth}{
\centering
\label{fig:douban_m2b_align}
\includegraphics[scale=0.5]{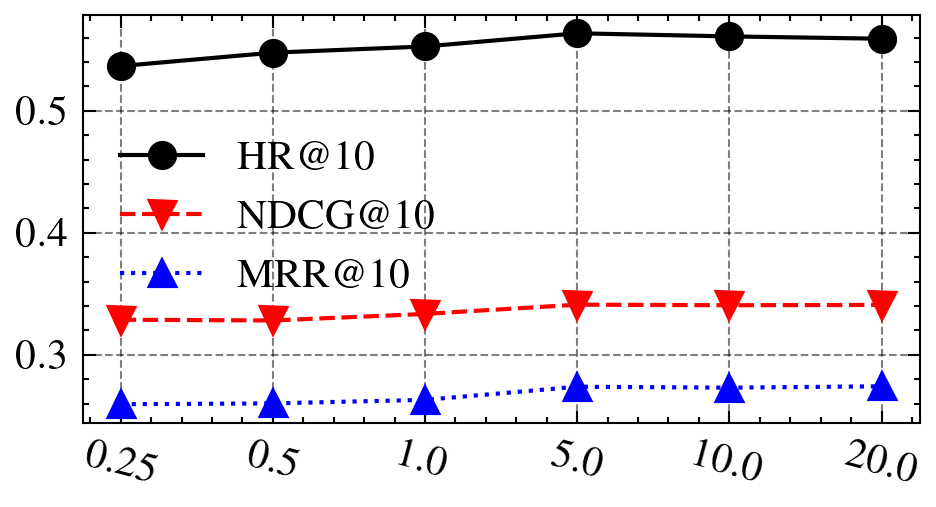} %
}
\end{minipage}
}
\subfigure[Effect of controller $\tau$]{
\begin{minipage}{0.47\columnwidth}{
\centering
\label{fig:douban_m2b_control}
\includegraphics[scale=0.5]{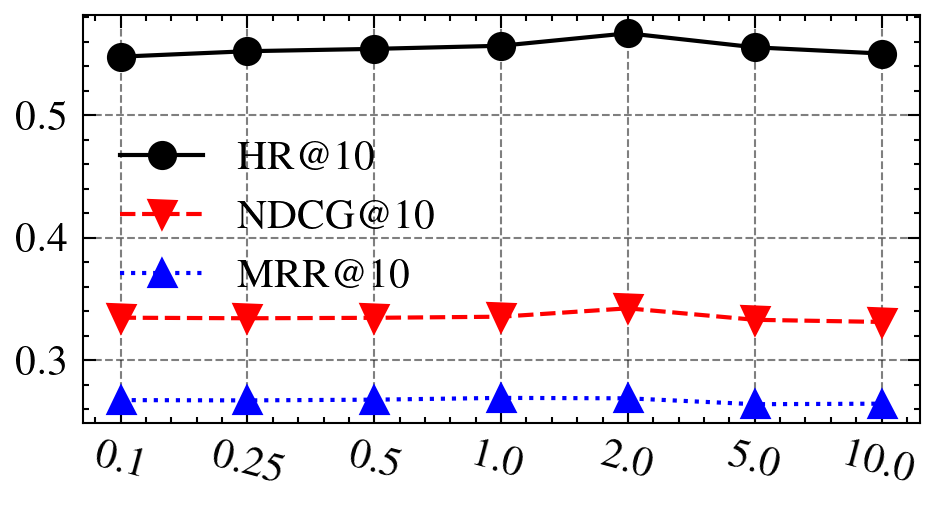} %
}
\end{minipage}
}

\subfigure[Effect of regularizing $\lambda_{R}$]{
\begin{minipage}{0.47\columnwidth}{
\centering
\label{fig:douban_m2b_rc}
\includegraphics[scale=0.5]{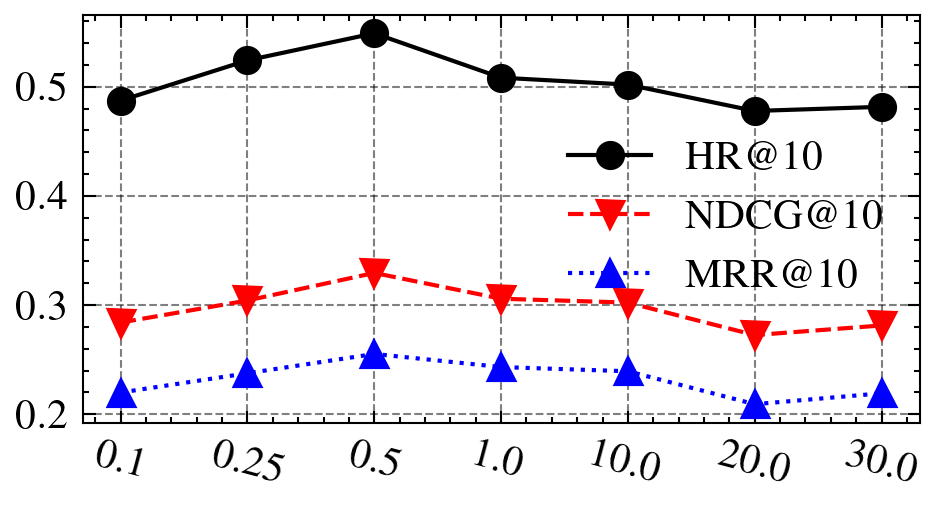} %
}
\end{minipage}
}
\subfigure[Effect of sparsity degree]{
\begin{minipage}{0.47\columnwidth}{
\centering
\label{fig:douban_m2b_sp}
\includegraphics[scale=0.48]{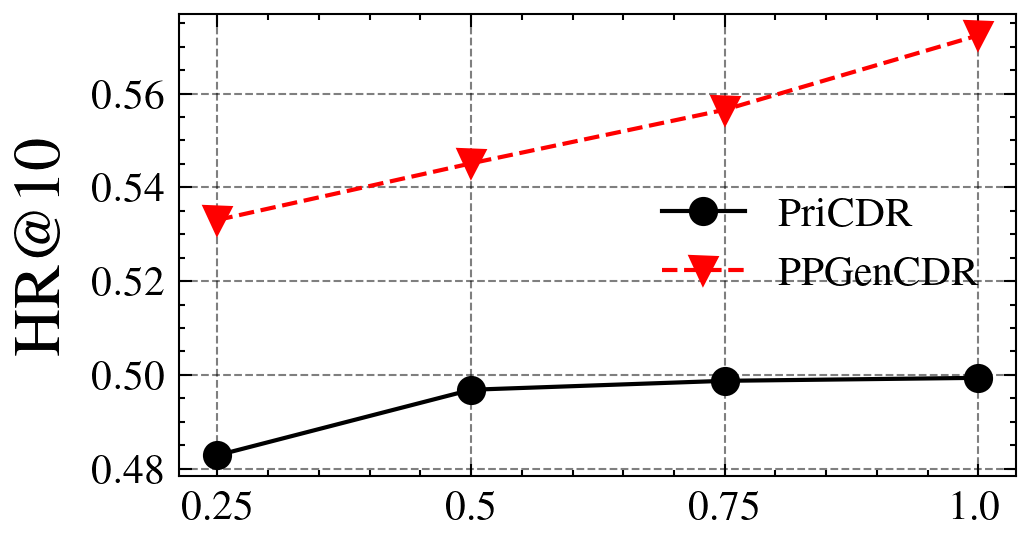} %
}
\end{minipage}
}
\caption{Parameters analysis on Douban movie $\rightarrow $ book.}
\label{fig:douban_m2b_sensitivity}
\end{figure}

\subsection{In-depth Model Analysis}
\nosection{Ablation Study (Q2)}
%
We remove GS and RC from \modelname~, to get \modelname-GS and \modelname-RC, in order to evaluate the utility of GS and RC, respectively.
%
From Tab.~\ref{tb:exp_res}, we can discover that: (1) 
%
Both \modelname-GS and \modelname-RC degrade their performance compared with \modelname.
This implies that the performance of PPCDR depends on not only stably modeling the distribution of private data in source domain, but also robustly using the perturbed information to recommend in the target domain. 
(2) 
Compared with \modelname-GS, the performance of \modelname-RC decreases more severely, meaning that it is indispensable to improve the robustness for PPCDR. 

\nosection{Empirical Study of Privacy (Q3)}
We analyze the effectiveness of SPP by replacing it with \modelname-SPP which exploits RDP to the gradients of \textit{the whole model}.
We evaluate the contribution of SPP from two aspects, i.e.,  (1) preserving the privacy of source domain and (2)
maintaining the performance of target domain, with two tasks, i.e., \textbf{Douban} movie $\rightarrow$ book, and \textbf{Amazon} movie $\rightarrow$ music.
Specifically, to study the former one, we compare the performance of the generators, published from 
\modelname~and \modelname-SPP, to
predict user preferences in the \textit{source} domain, and report the results in Tab.~\ref{tb:exp_res_s}. 
To evaluate the latter one, we depict the change of HR@10 in \textit{target} domain in Fig.~\ref{fig:privacy}, for \modelname~and \modelname-SPP, by varying privacy budget $\varepsilon \in \{ 0.04, 0.07, 0.3, 0.6, 1.2,2.4, 5, 11,16, 26, 72\}$. 
We can find that: (1) The generator of \modelname~leaks a little more privacy, since it has slightly better results than \modelname-SPP. 
(2) \modelname~significantly outperforms \modelname-SPP with different privacy budgets.
The results indicate that SPP is a cost-effective module for balancing preserving privacy and enhancing performance in PPCDR.

\begin{table}[t]
\renewcommand\arraystretch{.7}

\centering
\resizebox{\columnwidth}{!}{

{
\Huge
\begin{tabular}{lcccccc}

\toprule

              \multirow{1}{*}{}  & \multicolumn{3}{c}{\textbf{Douban} movie}             & \multicolumn{3}{c}{\textbf{Amazon} movie }            \\
              \midrule 
             & \textbf{HR}   & \textbf{NDCG} & \textbf{MRR}  &   \textbf{HR}   & \textbf{NDCG} & \textbf{MRR}   \\  
  \midrule
  \modelname & .0882 & .0258 & .0403 & .0945 & .0401 & .0522 \\
 \modelname-SPP & .0872 & .0201 & .0353 & .1098 & .0324 & .0500\\
 \bottomrule
\end{tabular}
}
}
\caption{Performance (top@10) of the generator of \modelname~in source domain. }
\label{tb:exp_res_s}
\end{table}

\nosection{Sensitivity of Hyper-parameters  (Q4)}
We show the effect of hyper-parameters on \textbf{Douban} movie $\rightarrow$ book in Fig.~\ref{fig:douban_m2b_sensitivity}.
In Fig.~\ref{fig:douban_m2b_sensitivity},
we compare the performance by varying the hyper-parameter of alignment $\lambda_{A} \in \{0.25,0.5,1,5,10,20\}$ in (a), the parameter of GS $\tau \in \{0.1, 0.25, 0.5, 1, 2, 5, 10\}$ in (b), and the hyper-parameter of robustness $\lambda_{R} \in \{0.1, 0.25, 0.5,1, 10, 20,30\}$ in (c), respectively. 
To study the effect of sparsity degree,
we compare the performance of \modelname~and PriCDR by randomly sampling 25\%, 50\%, 75\%, and 100\% of the data in source domain, and report the results in Fig.\ref{fig:douban_m2b_sensitivity} (d). 
From the results, we can conclude that: (1) \modelname~achieves the best performance when $\lambda_{A} = 1, \tau= 2$, and $\lambda_{R}= 0.5$. 
(2) The performance of \modelname~changes slightly with $\lambda_A$ in Fig.~\ref{fig:douban_m2b_sensitivity} (a), meaning that \modelname~can better leverage knowledge of the source domain by alignment while maintain the performance with RC. 
(3) 
The bell-shaped curves in Fig.~\ref{fig:douban_m2b_sensitivity} (b) and Fig.~\ref{fig:douban_m2b_sensitivity} (c) indicate that it is effective to choose a proper hyper-parameter for better recommendation performance. 
(4) \modelname~outperforms PriCDR under all sparsity degrees in Fig.~\ref{fig:douban_m2b_sensitivity} (d), indicating that \modelname~can better address the sparsity issue. 

\section{Conclusion} 
In this paper, we propose \modelname, i.e., a novel framework for privacy-preserving generative cross-domain recommendation. 
\modelname~devotes to stably modeling the distribution of private data in source domain by stable privacy-preserving
generator module, and robustly enhancing the performance of target domain by robust cross-domain recommendation module. 
The extensive empirical studies on two datasets (i.e., Douban and Amazon) demonstrate that \modelname~significantly outperforms the state-of-the-art recommendation models while preserving privacy.

\section{Acknowledgements}
This work was supported in part by the ``Pioneer" and ``Leading Goose" R\&D Program of Zhejiang (No. 2022C01126), the National Natural Science Foundation of China (No. 62172362), and Leading Expert of ``Ten Thousands Talent Program" of Zhejiang Province (No.2021R52001).

\bibliography{aaai23}
\appendix

\section{Experiments}
\label{sec:exp}
\subsection{Experimental Details}
\label{section:app_exp_details}
\noindent\textbf{Dataset statistics.} We report the statistics of two datasets, i.e., Douban ~\cite{zhu2021unified} and Amazon~\cite{ni2019justifying} in Tab.~\ref{tab:app_dataset}.

\noindent\textbf{Implementation details.}
For $\mathcal{G}$ in \moduleAname, we only input the user identification $\mathbf{u}_{id}$ to generate fake user preference, i.e., $\tilde{\boldsymbol{r}}_\mathcal{S} = \mathcal{G}(\mathbf{u}_{i d})$, following the existing work of GAN-based recommender systems~\cite{chae2018cfgan}. 
This is because our goal is to generate plausible user preference conditioned on the user identification $\mathbf{u}_{id}$ for each user correspondingly, instead of diversified and multiple user preferences for each user. 

\noindent\textbf{Evaluation methods.} Following \cite{chen2020towards}, we use the leave-one-out evaluation which randomly selects 2 rated items for validation and testing separately, and chooses 99 unrated items as negative samples for each user.

\begin{table}[h]
 \centering
    \caption{Datasets statistics of Amazon and Douban}
    \label{tab:app_dataset}
\begin{tabular}{cccccc}
\hline
\multicolumn{2}{c}{ \textbf{Datasets}}    &  \textbf{Users}  & \textbf{Items}  & \textbf{Ratings} & \textbf{Density} \\ \hline
\multirow{2}{*}{Amazon} & Music  & \multirow{2}{*}{13,460} & 18,467 & 218,346 & 0.088 \%  \\
                        & Book   &                        & 23,988 &  279,523 & 0.087 \%  \\ \hline
\multirow{2}{*}{Amazon} & Movie & \multirow{2}{*}{14,904} & 19,794 & 407,161 & 0.14\%  \\
                        & Music &                        & 20,058 & 278,545 & 0.09\%  \\ \hline
\multirow{2}{*}{Amazon} & Movie & \multirow{2}{*}{26704 } & 24091 & 562249 & 0.09\%  \\ 
                        & Book  &                        & 41884 & 574891 & 0.05\%  \\ \hline
\multirow{2}{*}{Douban} & Book  & \multirow{2}{*}{924}   & 3,916  & 50,429  & 1.39\%  \\
                        & Music &                        & 4,228 & 50,157 & 1.28\%  \\ \hline
\multirow{2}{*}{Douban} & Movie & \multirow{2}{*}{1574}  & 9,471  & 744,983  & 4.99\%  \\
                        & Book  &                        & 6,139 & 85,602 & 0.89\%  \\ \hline
\multirow{2}{*}{Douban} & Movie & \multirow{2}{*}{1055}  & 9,386 &  557,989 & 5.63\%  \\
                        & Music &                        & 4,981 & 60,626 & 1.15\%  \\ \hline
\end{tabular}
\end{table}

\section{Theoretical Analysis}
\label{sec:analysis}

\subsection{Privacy Analysis}

\begin{definition}[User-level Differential Privacy]
Let $\mathbf{R}$ be the user-item rating matrix, and $\boldsymbol{r}_i$ (the $i-${th} row of $\mathbf{R}$) be the preference vector of user $i$.
%
A mechanism $\mathcal{M}$ that takes the preference vector of user $i$, i.e., $\boldsymbol{r}_i$, as input, is $(\varepsilon, \delta)-$DP if for any differ-by-one user-item vectors $\boldsymbol{r}_{i}$, $\boldsymbol{r}^{\prime}_{i}$, and any event $\mathcal{A}$, the mechanism $\mathcal{M}$ satisfies:
\begin{equation}
\label{eq:app_udp}
\operatorname{P}[\mathcal{M}(\boldsymbol{r}_i) \in \mathcal{A}] \leq \exp (\varepsilon) \operatorname{P}\left[\mathcal{M}\left(\boldsymbol{r}^{\prime}_i\right) \in \mathcal{A}\right]+\delta,
\end{equation}
where $\varepsilon$ corresponds to privacy budget, and $\delta$ is the privacy uncertainty.
\end{definition}

\begin{definition}[\'Renyi Differential Privacy (RDP) \cite{mironov2017renyi}]
A mechanism $\mathcal{M}$ is $(\alpha, \varepsilon)-$RDP with order $\alpha \in (1, \infty)$ if for all neighboring datasets $\mathbf{S}$ and $\mathbf{S}^{\prime}$, the \'Renyi divergence satisfies: $D_{\alpha}\left(\mathcal{M}(\mathbf{S}) \| \mathcal{M}\left(\mathbf{S}^{\prime}\right)\right) \leq \varepsilon$, and further equals to $\left(\varepsilon+\frac{\log 1 / \delta}{\alpha-1}, \delta\right)-$DP, for any $\delta \in (0,1)$.  
\end{definition}

\begin{proposition}[RDP with Gaussian Mechanism \cite{dwork2014algorithmic,mironov2017renyi}]
\label{prop:gaussain_RDP}
For a $d$-dimensional function $f: X \rightarrow \mathcal{R}^{d}$ with sensitivity $\Delta_f= \max_{\mathbf{S}, \mathbf{S}^{\prime}} \|f(\mathbf{S}) - f(\mathbf{S}^{\prime})\|_{2}$, the Gaussian mechanism $\mathcal{M}_{\sigma_p} = f(x)+ \mathcal{N}(0,\sigma_p^2 \mathbf{I}) $, parameterized by the variance of Gaussian noise ${\sigma_p}$, is $(\alpha, \nicefrac{\alpha {\Delta_{f}}^2}{2 \sigma_p^2})-$RDP.
\end{proposition}

\begin{theorem}[Privacy Bound of Each \moduleAname ~Update Step]
\label{theorem:app_privacy_SPPG}
For batch size $N$, clipping constant $B$, and Gaussian mechanism $\mathcal{M}_{\sigma_{p}, B}$, each update step for training \moduleAname~satisfies
\begin{tiny}
$(\alpha, 2 N \alpha/\sigma_{p}^{2})-$RDP
\end{tiny} for $B=1$, and 
\begin{tiny}
$\left(2 N \alpha/\sigma_{p}^{2}+\frac{\log 1 / \delta}{\alpha-1}, \delta\right)-$DP. 
\end{tiny}
\begin{proof}
\noindent\textit{(1) (User-level \'Renyi Differential Privacy)} 
According to Proposition.~\ref{prop:gaussain_RDP}, for RDP with clip operation:
\begin{equation}
\label{eq:app_clip}
\operatorname{clip}(\boldsymbol{g}) =\max (1, \nicefrac{ \|\boldsymbol{g}\|}{B} ),
\end{equation}
we have the privacy sensitivity for user preference $\boldsymbol{r}$ with its differ-by-one neighbor $\boldsymbol{r}^{\prime}$ is:
\begin{equation}
    \Delta = max_{\boldsymbol{r}, \boldsymbol{r}^{\prime}} \|\mathcal{M}_{\sigma_{p}, B}(\boldsymbol{r}) - \mathcal{M}_{\sigma_{p}, B}(\boldsymbol{r}^{\prime}) \|_2 \leq 2B.
\end{equation}
Hence, we find that Gaussian mechanism, i.e., $\mathcal{M}_{\sigma_{p}, B} = \text{clip}(\boldsymbol{g}) + \mathcal{N}\left(0, \sigma_{p}^{2} B^{2} \boldsymbol{I}\right)$ is $(\alpha, 2\alpha/\sigma_{p}^{2})-$RDP, if $B=1$.

\noindent\textit{(2) (Composition)} For each update step of \moduleAname, we compute the sanitized gradients of discriminator $ \hat{\boldsymbol{g}}_{\mathcal{D}} $ and the original gradients of generator ${\boldsymbol{g}}_{\mathcal{G}}$, for $N$ user samples in a batch.
In this way, we only consider the privacy amplification in the discriminator:
\begin{small}
  \begin{equation}
   \label{eq:app_composition}
\begin{aligned}
\hat{\boldsymbol{g}}_{\mathcal{D}}&= \sum_{u=1}^N \mathcal{M}_{\sigma_{p}, B}(\nabla_{{\mathcal{D}}} \mathcal{L}_{\mathcal{D}}\left(\mu_{u} ;\boldsymbol{\theta}_{\mathcal{D}},\mathbf{u}_{u}\right)) \cdot J_{\boldsymbol{\theta}_{\mathcal{D}}},
\end{aligned}
\end{equation}
\end{small}
which can be treated as a composition of $N$ Gaussian mechanisms. 

\noindent\textit{(3) (Privacy Upper Bound of Batch Updating)} 
For the batched neighboring datasets $\mathbf{S} = \mathbf{R}_\mathcal{S}^{N \times N_{\mathcal{S}} }$ and $\mathbf{S}^{\prime} =\mathbf{\widetilde{R}}_\mathcal{S}^{N \times N_{\mathcal{S}} }$, we bound the \'Renyi divergence 
$D_{\alpha}\left(\hat{\boldsymbol{g}}_{\mathcal{D}}(\mathbf{S} ), \hat{\boldsymbol{g}}_{\mathcal{D}}\left(\mathbf{\mathbf{S}^{\prime} }\right)\right)$ to achieve a 
tighter upper bound.
Moreover, we can utilize the good properties of  \'Renyi divergence ~\cite{van2014renyi}, i.e., (a) data-processing inequality, and (b) additivity.

In the following, we denote $\mu$ and $\mu^{\prime}$ as the output distributions of applying Gaussian mechanism $\mathcal{M}_{\sigma_{p}, B}$ on $\boldsymbol{r}$ in $\mathbf{S}$ and  $\boldsymbol{r}^{\prime}$ in $\mathbf{S}^{\prime}$, respectively. 
We also denote $h_{i}(\cdot)$ as the post-processing function of the $i-$th user preference.
Therefore, we have the privacy bound of the batched neighboring datasets, which is measured by the \'Renyi divergence as follows:
\begin{small}
\begin{equation}
\begin{aligned}
&D_{\alpha}\left(\hat{\boldsymbol{g}}_{\mathcal{D}}(\mathbf{S}), \hat{\boldsymbol{g}}_{\mathcal{D}}\left(\mathbf{S}^{\prime}\right)\right) \\
&\leq D_{\alpha}\left(h_{1}\left(\mu_{1}\right) * \cdots * h_{N}\left(\mu_{N}\right) \| h_{1}\left(\mu_{1}^{\prime}\right) * \cdots * h_{N}\left(\mu_{N}^{\prime}\right)\right)  \text{ (5a)} \\
&\leq D_{\alpha}\left(\left(h_{1}\left(\mu_{1}\right), \cdots, h_{N}\left(\mu_{N}\right)\right) \|\left(h_{1}\left(\mu_{1}^{\prime}\right), \cdots, h_{N}\left(\mu_{N}^{\prime}\right)\right)\right) \text{ (5b)}  \\
&=\sum_{k} D_{\alpha}\left(h_{k}\left(\mu_{k}\right) \| h_{k}\left(\mu_{k}^{\prime}\right)\right) \text{ (5c)} \\
&\leq \sum_{k} D_{\alpha}\left(\mu_{k} \| \mu_{k}^{\prime}\right)  \text{ (5d)} \\
&\leq N \cdot \max _{k} D_{\alpha}\left(\mu_{k} \| \mu_{k}^{\prime}\right) \text{(5e)}  \\
&\leq N \cdot 2 \alpha / \sigma_{p}^{2} \quad \text{(5f)},
\end{aligned}
\end{equation}
\end{small}
where the sub-equations Eq.~(5a), Eq.~(5b), and Eq.~(5d), hold due to the data-processing inequality property of \'Renyi divergence, the sub-equation Eq.~(5c) holds since the additivity property of \'Renyi divergence, and the sub-equation Eq.~(5f) follows from \textit{(1) (User-level \'Renyi Differential Privacy)} in this proof. 

\noindent\textit{(4)} (\textit{Privacy Amplification}) Subsampling the dataset increases more privacy of a mechanism than directly using the whole dataset~\cite{wang2019subsampled}. 
We take the official implementation of RDP with sub-sampling implemented by~\cite{wang2019subsampled}.

\citet{wang2019subsampled} illustrate that if mechanism $\mathcal{M}_{\sigma_{p}, B}$ is
$(\alpha, \epsilon(\alpha))-$RDP for $\alpha \geq 1$, 
the subsampled mechanism $\operatorname{Subsample} \circ \mathcal{M}_{\sigma_{p}, B}$ follows $(\alpha, \epsilon^{\prime}(\alpha))-$RDP, where $\epsilon^{\prime}(\alpha)$ has the following lower bound:
\begin{tiny}
\begin{equation}
\begin{aligned}
\epsilon^{\prime}(\alpha) &\geq \frac{\alpha}{\alpha-1} \log (1-\gamma) \\
&+\frac{1}{\alpha-1} \log \left(1+\alpha \frac{\gamma}{1-\gamma}+\sum_{j=2}^{\alpha}\left(\begin{array}{c}
\alpha \\
j
\end{array}\right)\left(\frac{\gamma}{1-\gamma}\right)^{j} e^{(j-1) \epsilon(j)}\right),
\end{aligned}
\end{equation}
\end{tiny}
with $\gamma$ denoting the sampling rate.
\end{proof}
\end{theorem}

\subsection{Stability Analysis of the Backbone of \moduleAname}

\nosection{Dynamic Systems in Control Theory}
\label{section:app_clc}
In control theory, a \textit{signal} is a function that can be represented in both time domain and frequency domain.
And a \textit{dynamic system} is the way that a signal (e.g., $\boldsymbol{m}(t)$) derives another one (e.g., $\boldsymbol{y}(t)$), over time.
The dynamic system can be formulated by differential equation:
\begin{equation}
\label{eq:DE}
\frac{d \boldsymbol{y}(t)}{d t}=f(\boldsymbol{y}(t), \boldsymbol{m}(t)),
\end{equation}
with an initial condition $\boldsymbol{y}(0) =\boldsymbol{y}_0$.
We denote signals in time domain by bold lowercase letters (e.g., $\boldsymbol{y}$, $\boldsymbol{m}$), and signals in frequency domain by bold uppercase letters (e.g., $\boldsymbol{Y}$, $\boldsymbol{M}$) in this paper.
%
%
 $\boldsymbol{Y}(s)=\boldsymbol{T}(s) \boldsymbol{M}(s)$ represents a dynamic system like Eq.~\eqref{eq:DE} in frequency domain, where $s$ is the complex number frequency parameter, and $\boldsymbol{T}(s)$ is a transfer function.

Control theory provides a series of methods to analyze and improve the stability, which are based on the the poles of transfer function.

To \textbf{analyze the stability} of a dynamic system, we have a property as follows.
\begin{proposition}[Stability of dynamic system \cite{kailath1980linear}]
\label{prop:stability}
A dynamic system is asymptotic stable, if all the poles of its transfer function have negative real parts. 
\end{proposition} 

To \textbf{improve the stability} of a dynamic system, closed-loop control (CLC) system~\cite{kailath1980linear} is widely used for a non-linear dynamic system.
Specifically, CLC system has an additional controller to adjust transfer function between the output signals and input signals, respectively, given the feedback of output.

%



\begin{theorem}[Stability Analysis of the Backbone of \moduleAname]
\label{theorem:app_stability}
The modeling process of \moduleAname~is unstable when using Gaussian mechanism  $\mathcal{M}_{\sigma_{p}, B}$ with noise $\sigma_{\text{gr}} \sim \mathcal{N}\left(0, \sigma_{p}^{2} B^{2} \boldsymbol{I}\right)$ to preserve privacy, and the stability of the $\mathcal{G}$ depends on the stability of the $\mathcal{D}$.
\begin{proof}

In this theorem, we analyze the stability of \textbf{Eq.~(5) in our main paper}. 
Firstly, we formulate \moduleAname~as a dynamic system, which can be simplified by Dirac GAN for stability analysis.
Next, we take the effect of perturbation brought by selective privacy preserver (SPP in our main paper), into the dynamic system of \moduleAname.
Then, we obtain the transfer function of the dynamic system of \moduleAname~in the frequency domain.
Lastly, we analyze the stability based on the control theory.

\noindent\textit{(1) (Formulation)} We first simplify \moduleAname~to a Dirac GAN~\cite{mescheder2018training}, which is widely studied to analyze the stability of GAN-based models.
In Dirac GAN, the ground truth distribution of $\boldsymbol{r}$ is $p(\boldsymbol{r})=\text{Dirac}(\boldsymbol{r}-\boldsymbol{c})$, where all elements of the vector $\boldsymbol{c}$ are constant.
And the generator $\mathcal{G}$ models a distribution ${p_{\mathcal{G}}(\boldsymbol{r})} = \text{Dirac} (\boldsymbol{r}- \boldsymbol{\theta}_{\mathcal{G}})$, and the discriminator $\mathcal{D}$ is defined as $\mathcal{D}(\boldsymbol{r})= \boldsymbol{\theta}_{\mathcal{D}} \boldsymbol{r}$. 
Optimizing a Dirac GAN with the objective of \moduleAname, we have the gradient step as follows:
\begin{small}
\begin{equation}
\label{eq:app_g_d_s}
\left\{\begin{aligned}
\frac{d \boldsymbol{\theta}_{\mathcal{D}}}{d t} &=\boldsymbol{c} -\boldsymbol{\theta}_{\mathcal{G}}(t), \\
\frac{d \boldsymbol{\theta}_{\mathcal{G}}}{d t} &=\boldsymbol{\theta}_{\mathcal{D}}(t).
\end{aligned}\right.
\end{equation}
\end{small}

\noindent\textit{(2) (Formulation with Privacy-preserving)} Note that \moduleAname~uses SPP to apply Gaussian Mechanism with noise $\sigma_{\text{gr}} $ range from $[-\sigma_{p} B, \sigma_{p} B]$ on the gradients, thus the ground truth distribution is not definite  during the modeling procedure. 
To simulate it, we substitute the ground truth distribution of each user preference $\boldsymbol{c}$ in $p(\boldsymbol{r})$ with $\boldsymbol{m}(t) = \boldsymbol{c} -(1-2 \phi)\boldsymbol{\sigma}_{\text{gr}}$ $(\phi \in [0,1])$
by interpolating the noise brought from Gaussian mechanism.

\noindent\textit{(3) (Frequency Domain Transformation)} Thus we have the optimization of \moduleAname~with a distribution for the perturbed ground truth, and transform it to frequency domain as follows:
\begin{small}
\begin{equation}
\label{eq:app_dy_T2F}
\begin{aligned}
\left\{\begin{array} { r l } 
\frac{d \boldsymbol{\theta}_{\mathcal{D}}}{d t}&=\boldsymbol{m}(t)-\boldsymbol{\theta}_\mathcal{G} (t), \\
\frac{d \boldsymbol{\theta}_{\mathcal{G}}}{d t} &= \boldsymbol{\theta}_\mathcal{D} (t),
\end{array} 
\Rightarrow 
\left\{\begin{array}{l}
\boldsymbol{\Theta}_\mathcal{D}(s)=\frac{s}{s^{2}+1} \boldsymbol{M}(s), \\
\boldsymbol{\Theta}_\mathcal{G}(s)=\frac{1}{s} \boldsymbol{\Theta}_\mathcal{D}(s)=\frac{1}{s^{2}+1} \boldsymbol{M}(s).
\end{array}\right.\right.
\end{aligned}
\end{equation}
\end{small}
Let the output signal $\boldsymbol{y}(t)=(\boldsymbol{\theta}_\mathcal{D}( t),\boldsymbol{\theta}_\mathcal{G}(t))$, $\forall t>0$.
And we denote the transfer function of the $\mathcal{D}$ as $\boldsymbol{T}_\mathcal{D}(s) = \nicefrac{s}{(s^2+1)}$ and the transfer function of the $\mathcal{G}$ as $\boldsymbol{T}_\mathcal{G}(s) = \nicefrac{1}{(s^2+1)}$ respectively.
The left arrow in Eq.~\eqref{eq:app_dy_T2F} holds by applying Laplace transform on the both sides of equation.

\noindent\textit{(vi) (Conclusions)} The dynamic system in Eq.~\eqref{eq:app_dy_T2F} is \textit{not stable} from Proposition.~\ref{prop:stability}, because both the poles of transfer functions in $\mathcal{D}$ and $\mathcal{G}$ are exactly on the complex axis (not negative real parts).
Additionally, in terms of Eq.~\eqref{eq:app_dy_T2F}, the updating of  $\mathcal{G}$ in frequency domain depends on  $\mathcal{D}$.
That is, the stability of \moduleAname~is controlled once $\mathcal{D}$ is stable.
\end{proof}
\end{theorem}

\subsection{Stability of Using GAN Stabilizer in \moduleAname}
\begin{theorem}[Stability of Using GAN Stabilizer in \moduleAname~(Eq.~(6) in our main paper)]
\label{theorem:app_gs_control}
GAN stabilizer (GS) parameterized by $\tau (\tau > 0)$, i.e., 
$\boldsymbol{T}_{GS}(s)=\tau (\tau > 0)$,
is a controller to make the dynamic system of $\mathcal{D}$ stable.
\begin{proof}

In this theorem, we prove that the \textbf{Eq.~(6) in our main paper} is stable based on the control theory mentioned at the beginning of Appendix B.2.
Additionally, we explain for the induction of the regularization term of GAN stabilizer, which is \textbf{$\mathcal{L}_{\text{GS}}$ in our main paper}.
Given the dynamic system of updating step for \moduleAname~with GAN stabilizer in time domain:
\begin{equation}
\label{eq:app_update_step}
\left\{\begin{aligned}
\frac{d \boldsymbol{\theta}_{\mathcal{D}}}{d t} &=\boldsymbol{m}(t)-\boldsymbol{\theta}_{\mathcal{G}}(t) -\tau \boldsymbol{\theta}_{\mathcal{D}}(t), \\
\frac{d \boldsymbol{\theta}_{\mathcal{G}}}{d t} &=\boldsymbol{\theta}_{\mathcal{D}}(t),
\end{aligned}\right.
\end{equation}
we use Laplace Transform~\cite{widder2015laplace} to convert it into the frequency domain:
\begin{equation}
\label{eq:app_clc_FT}
\left\{\begin{array}{l}
\boldsymbol{\Theta}_{\mathcal{D}}(s)=\frac{s}{s^{2}+\tau s+1} \boldsymbol{M}(s), \\
\boldsymbol{\Theta}_{\mathcal{G}}(s)= \frac{1}{s}\boldsymbol{\Theta}_{\mathcal{D}}(s) =\frac{1}{s^{2}+\tau s+1} \boldsymbol{M}(s),
\end{array}\right.
\end{equation}
where $\boldsymbol{\Theta}_\mathcal{D}(s)$, $\boldsymbol{\Theta}_\mathcal{D}(s)$, and $\boldsymbol{M}_(s)$ are the notations of $\boldsymbol{\theta}_\mathcal{D}(s)$, $\boldsymbol{\theta}_\mathcal{D}(s)$, and $\boldsymbol{m}_(s)$, respectively, in frequency domain.

In Eq.~\eqref{eq:app_clc_FT}, both the complex roots of $s^2 +\tau s +1$ ($\tau >0$) are in the negative real parts.
Thus, compared with~Eq.~\eqref{eq:app_dy_T2F}, the stability of \moduleAname~is controlled by the extra controller, i.e., GS, from the view of Proposition~\ref{prop:stability}.

By now, we have proved the stability of updating \moduleAname~with GS.
Next, we will provide the additional explanation for the regularization term, i.e., $\mathcal{L}_{GS}$, in the objective of optimizing \moduleAname. 

We first integrate the updating step, i.e., Eq.~\eqref{eq:app_update_step}, to get the objective of optimization to maximize for \moduleAname~as follows:
\begin{small}
\begin{equation}
\label{eq:dg}
\left\{
\begin{aligned}
\mathbf{L}_\mathcal{G} &= \mathbb{E}_{\tilde{\boldsymbol{r}} \sim p_{\tilde{\boldsymbol{r}}_\mathcal{S}})}\left[\mathcal{D}\left(\tilde{\boldsymbol{r}},\mathbf{u}_{i d}\right)\right],\\
\mathbf{L}_\mathcal{D} &= \mathbb{E}_{\boldsymbol{r} \sim p_{\boldsymbol{r}_\mathcal{S}}}\left[\mathcal{D}\left(\boldsymbol{r}, \mathbf{u}_{i d}\right)\right] - \mathbb{E}_{\tilde{\boldsymbol{r}} \sim p_{\tilde{\boldsymbol{r}}_\mathcal{S}}}\left[\mathcal{D}\left(\tilde{\boldsymbol{r}},\mathbf{u}_{i d}\right)\right] \\
& - \frac{\tau}{2}\left(\mathbb{E}_{\boldsymbol{r} \sim  p_{\boldsymbol{r}_\mathcal{S}} }\left[\mathcal{D}^{2}\left(\boldsymbol{r}, \mathbf{u}_{i d}\right)\right]+\mathbb{E}_{{\tilde{\boldsymbol{r}} \sim p_{\tilde{\boldsymbol{r}}_\mathcal{S}}} }\left[\mathcal{D}^{2}\left( \tilde{\boldsymbol{r}}, \mathbf{u}_{i d}\right)\right]\right),
\end{aligned}
\right.
\end{equation}
\end{small}
which exactly equals to the objective of optimization to maximize in our main paper:
\begin{small}
\begin{equation}\nonumber
\left\{
\begin{aligned}
&\mathbf{L}_\mathcal{G} = \mathcal{L}_\mathcal{G},\\
&\mathbf{L}_\mathcal{D} = \mathcal{L}_\mathcal{D} +  \mathcal{L}_{\text{GS}}.
\end{aligned}
\right.
\end{equation}
\end{small}

To conclude, we first prove the effectiveness of using GS to stabilize \moduleAname, which uses SPP to preserve the privacy of interaction data in the source domain. 
Then, we explain how $\mathcal{L}_{GS}$ becomes an extra regularization term in the objective of optimizing \moduleAname. Here ends our proof.
\end{proof}
\end{theorem}

\end{document}